\documentclass[journal=jacsat,manuscript=article]{achemso}
\SectionNumbersOn
\usepackage{enumitem}
\usepackage{xspace}
\usepackage{fancyhdr}
\usepackage{afterpage}
\usepackage[usenames,dvipsnames]{xcolor}
\usepackage{dcolumn}
\usepackage{comment}
\usepackage{listings}
\usepackage{makeidx}
\usepackage{longtable}
\usepackage{chemformula}
\usepackage[utf8]{inputenc}
\usepackage{comment}
\usepackage[dvips]{epsfig}
\usepackage{epstopdf}
\usepackage{graphicx}
\usepackage{booktabs}
\usepackage{mathtools}
\usepackage{amsmath}
\usepackage{amssymb}
\usepackage[version=4]{mhchem}
\usepackage{amsfonts}
\usepackage{commath} 
\usepackage{gensymb}
\usepackage{adjustbox}
\usepackage{subfiles}
\usepackage{xr}
\newcommand{\kalpha}{\ensuremath{K_\alpha}\xspace}

\newcommand{\qmax}{\ensuremath{Q_{\mathrm{max}}}\xspace}

\newcommand{\qmin}{\ensuremath{Q_{\mathrm{min}}}\xspace}
\newcommand{\qbroad}{\ensuremath{Q_{\mathrm{broad}}}\xspace}
\newcommand{\qdamp}{\ensuremath{Q_{\mathrm{damp}}}\xspace}

\newcommand{\uiso}{\ensuremath{U_{\mathrm{iso}}}\xspace}

\newcommand{\rpoly}{\ensuremath{rpoly}\xspace}
\newcommand{\rw}{\ensuremath{R_w}\xspace}

\newcommand{\fq}{\ensuremath{F(Q)}\xspace}
\newcommand{\gr}{\ensuremath{G(r)}\xspace}

\newcommand{\q}{\ensuremath{Q}\xspace}

\newcommand{\tth}{\ensuremath{2\theta}\xspace}

\renewcommand{\vec}[1]{\mathbf{#1}}
\newcommand{\iaa}{\AA\ensuremath{^{-1}}\xspace}

\newcommand{\sjba}[1]{}

\newcommand{\mta}[1]{}

\newcommand{\eq}[1]{Eq.~\ref{eq:#1}}

\newcommand{\fig}[1]{Fig.~\ref{fig:#1}}

\newcommand{\adhoc}{\textit{ad hoc}\xspace}

\graphicspath{ {./figures/} }
\definecolor{dgreen}{HTML}{008000}

\newcommand{\getx}{{\sc PDFgetX3}\xspace}

\newcommand{\cmi}{{\sc Diffpy-CMI}\xspace}

\newcommand{\cve}{\ensuremath{c_{v_e}}\xspace}   
\newcommand{\mud}{\ensuremath{\mu D}\xspace}
\newcommand{\mude}{\ensuremath{\mu D_e}\xspace}
\newcommand{\mudmone}{\ensuremath{\mu D_{m1}}\xspace}
\newcommand{\mudmtwo}{\ensuremath{\mu D_{m2}}\xspace}
\newcommand{\mudmthree}{\ensuremath{\mu D_{m3}}\xspace}
\newcommand{\mudmfour}{\ensuremath{\mu D_{th}}\xspace}
\newcommand{\labpdfproc}{\textsc{diffpy.labpdfproc}\xspace}
\newcommand{\diffpy}{\textsc{diffpy}\xspace}
\newcommand{\utils}{\textsc{diffpy.utils}\xspace}
   
\newcommand{\zscan}{$z$-scan\xspace}
\newcommand{\zscans}{$z$-scans\xspace}
\newcommand{\citexraydb}{\cite{newvilleXrayDBXrayReference2024}}

\author{Yucong Chen}
\affiliation{Department of Applied Physics and Applied Mathematics, Columbia University, New York, NY~10025, USA}
\altaffiliation{Y.C. and T.S. contributed equally to this work}
\author{Till Schertenleib}
\affiliation{Institute of Chemical Sciences and Engineering (ISIC), École Polytechnique Fédérale de Lausanne (EPFL), Sion, CH-1951, Switzerland}
\altaffiliation{Y.C. and T.S. contributed equally to this work}
\author{Andrew Yang}
\affiliation{Department of Applied Physics and Applied Mathematics, Columbia University, New York, NY~10025, USA}
\author{Pascal Schouwink}
\affiliation{Institute of Chemical Sciences and Engineering (ISIC), École Polytechnique Fédérale de Lausanne (EPFL), Sion, CH-1951, Switzerland}
\author{Wendy~L. Queen}
\affiliation{Institute of Chemical Sciences and Engineering (ISIC), École Polytechnique Fédérale de Lausanne (EPFL), Sion, CH-1951, Switzerland}
\author{Simon~J.~L. Billinge}
\email{sb2896@columbia.edu}
\affiliation{Department of Applied Physics and Applied Mathematics, Columbia University, New York, NY~10025, USA}

\title{An Absorption Correction for Reliable Pair-Distribution Functions from Low Energy X-ray Sources}

\begin{document}


\begin{abstract}
This paper explores the development and testing of a simple absorption correction model for processing x-ray powder diffraction data from Debye-Scherrer geometry laboratory x-ray experiments. 
This may be used as a pre-processing step before using \getx to obtain reliable pair distribution functions (PDFs).
The correction was found to depend only on \mud, the product of the x-ray attenuation coefficient and capillary diameter.
Various experimental and theoretical methods for estimating \mud were explored, and the most appropriate \mud values for correction were identified for different capillary diameters and x-ray beam sizes.
We identify operational ranges of \mud where reasonable signal to noise is possible after correction.
A user-friendly software package, \labpdfproc, is presented that can help estimate \mud and perform absorption corrections, with a rapid calculation for efficient processing.
\end{abstract}

\section{Introduction}

Historically, the study of local atomic structures using x-ray pair distribution function (PDF) analysis has predominantly been done on data from synchrotron facilities using rapid acquisition PDF (RAPDF) techniques \cite{chupa;iucrcpd03}. 
However, using laboratory-based x-ray sources to get good PDFs is of great interest.
This has been demonstrated from laboratory diffractomers equipped with Mo and Ag $K_\alpha$ sources in various examples including crystalline materials \cite{confalonieriComparisonTotalScattering2015} and nanoparticles \cite{prinzHardXraybasedTechniques2020}.

The current authors recently tested protocols for optimizing experimental setups for such measurements \cite{schertenleibProtocolsObtainingReliable2024}.
That work discussed that some factors affecting the data reduction to obtain quantitatively accurate PDFs that are often neglected in RAPDF experiments become more relevant for lab-based experiments, especially for those done with Mo~\kalpha radiation.
In particular, multiplicative corrections to the measured intensities from sample absorption are likely to have a non-negligible \q-dependence for Mo~\kalpha data but not for RAPDF data \cite{egami;b;utbp12,schertenleibProtocolsObtainingReliable2024}.
This is for two reasons.
First, the lower x-ray energies result in overall higher sample absorption effects.
Second, in lab experiments it is necessary to measure over a large range of \tth, typically up to 140\degree~or higher with Mo or Ag x-rays, making the intensities more susceptible to any angle-dependent multiplicative corrections.

The influence of sample absorption has been discussed for powder diffraction in general. 
Absorption leads to differences in calculated and measured powder diffraction patterns and has to be accounted for in Rietveld refinements \cite{pitschkeIncorporationMicroabsorptionCorrections1993}. 
In Bragg-Brentano geometries finite sample thickness effects must be considered when measuring low absorbing specimens \cite{dinnebierRietveldRefinementPractical2018,egami;b;utbp12}.
Absorption effects are also crucial in successful quantitative XRD analysis by multiphase refinements \cite{brindleyXLVEffectGrain1945, taylorAbsorptionContrastEffects1991, lippExtensionRietveldRefinement2022}. 
Here, we present a detailed investigation of the influence of sample absorption in bench-top PDF analysis, i.e., for lab diffractometers with capillary (Debye-Scherrer) geometry.
We present a software package for making sample-absorption corrections to measured data for this geometry.

Among the numerous available software packages to obtain \gr from raw data \cite{petkovRADProgramAnalysis1989, jeongPDFgetXProgramDetermining2001, qiuPDFgetX2GUIdrivenProgram2004, soper2011gudrunn}, \getx \cite{juhasPDFgetX3RapidHighly2013} is widely used by the community as it follows a simple \adhoc approach to data reduction \cite{billingeRobustAdHoc2013}. 
The \adhoc algorithm used in \getx does a good job of correcting for parasitic scattering (unwanted additive contributions to the signal) but makes no correction for multiplicative effects. 
This works well for RAPDF data where they are small and their angle dependence is even smaller.
However as discussed above, when applied to data from a laboratory diffractometer this simplifying assumption may not be valid in general, which motivated this study.

This work also allows us to suggest best practices for sample preparation for PDF experiments on laboratory x-ray diffractometers to mitigate the worst effects of sample absorption.  
These strategies are not new, but here we validate and quantify them somewhat to give more precise guidance.

\section{Absorption correction}
\label{sec:abs_corr}

Here, we address the multiplicative correction due to sample self-absorption effects for the most common geometry used for PDF analysis, the Debye-Scherrer geometry, in which a beam is incident on a cylindrical sample in a capillary with a detector rotating around the sample.
The absorption correction for this geometry was developed in detail many years ago \cite{paalmanNumericalEvaluationXRay1962, kendigXRayAbsorptionFactors1965, poncettInelasticScatteringLiquid1977, soperMultipleScatteringAttenuation1980} and is routinely applied in neutron scattering measurements.  
Various codes implement these corrections as part of the total scattering data reduction workflow\cite{qiuPDFgetX2GUIdrivenProgram2004,tobyGSASIIGenesisModern2013,soperExtractingPairDistribution2011}.
They require rather detailed knowledge of the sample composition and packing fraction, and for the greatest accuracy, also the sample and sample-container geometry.

In this work, we are more concerned with obtaining sufficient accuracy in these corrections in combination with \adhoc data reduction approaches used as \getx, and understanding the effects of approximate corrections on resulting PDFs.
We explore this for a range of possible sample compositions with Mo~\kalpha and Ag~\kalpha radiation.
We first briefly describe a simplified derivation of the angle dependence of the absorption in this geometry and then explore the nature of the resulting curves in different situations.
The full derivation is in the Supplementary Information, but is summarized here as it can help to build intuition about absorption effects in a laboratory x-ray setting.

\begin{figure}
    \centering
    \includegraphics[width=0.8 \textwidth]{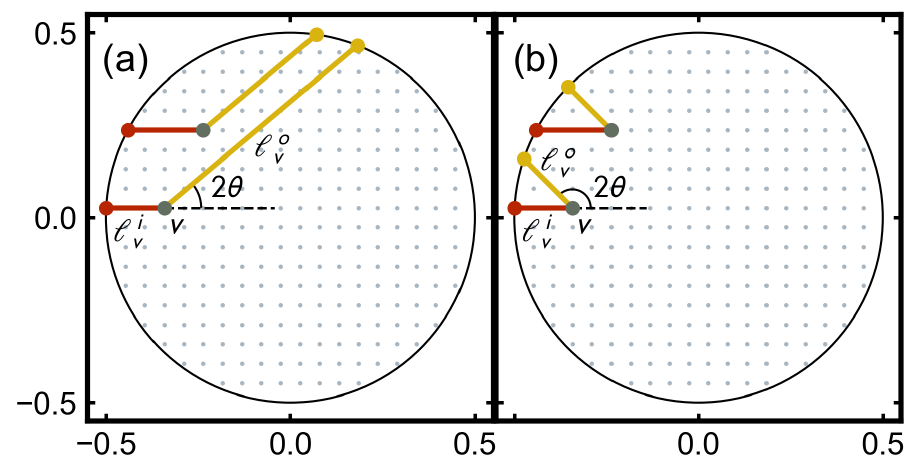}
    \caption{
    Incoming and outgoing x-ray beam paths for x-rays undergoing scattering in two representative pixels at an angle (a) $\tth = 40 \degree$ and (b) $\tth = 135 \degree$. In each case, the large black circle indicates the edge of a cross-section of the cylindrical sample capillary. The x-rays arrive horizontally from the left. A scattered ray is then shown at the given \tth angle. The small blue-grey dots indicate positions of a uniform grid of voxels in the circle and the red and yellow dots indicate the point on the surface of the sample where the x-rays enter and exit, respectively. The path-length for the x-ray scattered in the $v$th voxel at angle \tth is then $\ell_v(\tth) = \ell_v^i + \ell_v^o(\tth)$, the sum of the lengths of the red line and the yellow lines.}
    \label{fig:path_lengths} 
\end{figure}

Assuming a homogeneous ideal powder in the absence of sample absorption the measured x-ray intensity at some point in $\q$, $I_m(\q)$, would be proportional to the illuminated sample volume, $V$.  Here $\q$ is the magnitude of the scattering vector, $\q = |\vec{k}_i - \Vec{k}_s| = \frac{4\pi \sin\theta}{\lambda}$ where $\vec{k}_i$ and $\vec{k}_s$ are the incident and scattered wave vectors, respectively, $\theta$ is the Bragg angle which is half the scattering angle, \tth, which is the angle between the incident and scattered x-ray beams (see \fig{path_lengths}) and $\lambda$ is the x-ray wavelength.
A normalized coherent scattering intensity per unit volume of sample, $i_c$, can therefore be obtained by dividing $I_m$ by $V$.

In the presence of sample absorption, the scattered intensity from any small volume element (voxel) in the sample will be reduced by absorption of the x-ray as it travels along the incoming, $\ell_v^i$, and outgoing, $\ell_v^o$, path through the sample.
This is shown in \fig{path_lengths} for the case of a cylindrical capillary mounted perpendicular to the beam.
The derivation of the path lengths is reproduced in detail in Section~\ref{sec:path_length_si} in the Supplementary Information.
When there is significant sample absorption, to get $i_c$ we would divide $I_m$ not by the full illuminated volume but by an absorption corrected effective volume, $V_e$, so that
\begin{equation}
    i_c (2 \theta) = \frac{I_m (2 \theta)}{V_e (2 \theta)},
    \label{eq:eff_vol_norm}
\end{equation}
where the effective volume is given by \cite{schertenleibProtocolsObtainingReliable2024},
\begin{equation}
    V_e (2 \theta) = \sum_v e^{-\mu_s^\lambda\ell_v(\tth)} \Delta V_v,
\end{equation}
where $\Delta V_v$ is the volume of the $v$th voxel and $\mu_s^\lambda$ is the linear absorption coefficient of the sample for x-rays of wavelength $\lambda$. $\ell_v(\tth)$ is the total path-length, i.e., the sum of the incident and outgoing path lengths (see \fig{path_lengths}), 
\begin{equation}
\ell_v(\tth) = \ell_v^i+\ell_v^o(\tth),\label{eq:ell_tth}
\end{equation}
for the $v$th voxel and scattering angle \tth.
The normalized intensity from \eq{eff_vol_norm} is then
\begin{equation}
    i_c = \frac{I_m}{\sum_v e^{-\mu_s^\lambda\ell_v}\Delta V_v}.
\end{equation}
For the case where all the voxels have the same volume $\Delta V_v = V/N_v$ and we get
\begin{equation}
    i_c = \frac{N_v I_m}{V\sum_v e^{-\mu_s^\lambda\ell_v(\tth)}}\label{eq:ic-final}.
\end{equation}
We can then define an absorption correction $c_{v_e}$ as
\begin{equation}
    c_{v_e}(\tth) = \frac{N_v}{\sum_v e^{-\mu_s^\lambda\ell_v(\tth)}}.
\end{equation}

Both the strength of the x-ray attenuation, and its angle dependence, are dependent on the material and wavelength specific linear absorption coefficient, $\mu_s^\lambda$.
It also depends on a sum over all the path lengths an x-ray takes through the sample as it arrives from the source and exits after scattering.

For simpler flat geometries it is known that the curve shape depends only on the product $\mu t$, where $t$ is the thickness of the sample, and not independently on $\mu$ and $t$. 
We show here that this is also true for the capillary geometry: the curve depends only on the product \mud where $D$ is the capillary diameter.
This greatly simplifies the analysis by reducing the dimensionality of the space of possibilities we need to consider.
Details of the proof are provided in Section~\ref{sec:mud_si} in the Supplementary Information.
There we show that \mud can be factored out of the sum, $\sum_v e^{-\mu_s^\lambda\ell_v(\tth)}$ in \eq{correction-term} which can be rewritten as
\begin{equation}
    c_{v_e}(\tth) = \frac{N_v}{\sum_v e^{-\mu_s^\lambda D \cdot d\ell_v(\tth)}},
    \label{eq:correction-term}
\end{equation}
where $d\ell_v(\tth)$ is the path length that the x-ray would traverse for a capillary of unit diameter.
Thus, the absorption correction depends not on $\mu$ and $D$ independently but on the product \mud (where for compactness, we drop hereafter the superscript and subscript that explicitly indicate that it is \mud for the sample at a particular wavelength).

We explore below the effect on the PDF of applying this correction to data from a variety of samples of different absorptions, discuss different ways of estimating \mud for a given sample, and describe a software package for rapidly obtaining and applying the correction.

\section{Validation Experiments}

To explore the effects of the correction on real data, powder x-ray diffraction patterns were collected for four samples with varying absorption cross-sections. To cover a large range of \mud, \ce{ZrO2}, \ce{CeO2}, and \ce{HfO2} were packed in Kapton (polyimide tubes) with varying inner diameters (ID's).
To get a sense of how the \mud's are distributed, the theoretical values were computed based on each sample composition, mass density, and capillary diameter using the XrayDB database~\citexraydb. The mass density was determined by measuring the mass of the packed powder and the length of the powder bed.
The sample list, along with the corresponding information, are presented in Table \ref{tab:exp_list}, ranging from $\mud = 0.8$ to almost 12.
For reference, a subset of the samples (ID=1 mm) were also measured using synchrotron x-rays.

\begin{table}
    \caption{List of samples that were measured on the Bruker lab diffractometers. The samples were packed in Kapton (polyimide) tubes with different inner diameters (ID). The density was calculated based on the amount of powder packed in a given segment of the cylindrical Kapton tubes. \mud's were calculated using the database from XrayDB.}
    \begin{center}
    \begin{tabular}{lccc}
    \toprule
    sample & ID (mm) & density ($g/cm^{3}$) & \mud \\
    \midrule
    \ce{ZrO2} & 635 & 1.009 & 0.795 \\
    ~ & 813 & 0.856 & 0.864 \\
    ~ & 1024 & 1.122 & 1.426 \\
    \ce{CeO2} & 635 & 1.706 & 4.229 \\
    ~ & 813 & 1.435 & 4.554 \\
    ~ & 1024 & 1.457 & 5.824 \\
    \ce{HfO2} & 635 & 1.741 & 8.168 \\
    ~ & 813 & 1.963 & 11.79 \\
    ~ & Wire$^a$ & -- & -- \\
    \bottomrule
    \end{tabular}
    \label{tab:exp_list}
    \end{center}
    $^a$powder was rubbed on the outside of a glass wire covered in grease. The density and diameter of wire and powder are therefore unknown, and a theoretical \mud cannot be computed.
\end{table}


The laboratory PDF measurements were performed on a Bruker~D8 Discovery diffractometer equipped with a Mo~\kalpha source ($K_{\alpha_1\alpha_2}$ double emission, average wavelength $\lambda$=0.71073~\AA) using a capillary geometry to ensure a constant sample illumination.
The configuration included a focusing Goebel mirror, a divergence slit of 1.0~mm for IDs 0.635~mm and 0.813~mm, and 1.2~mm for IDs 1.024~mm, a~2.5\degree~axial Soller slit, a scattering guard after the source, and an additional~2.5\degree~axial Soller slit in the diffraction beam before the detector. 
X-ray generator settings of 50~kV and 50~mA were employed. 
The acquisitions were conducted using the staircase-counting-time (SCT) measurement strategy described in our previous work that ensures increased counting statistics in the high-\q region\cite{schertenleibProtocolsObtainingReliable2024}.

The SCT acquisition protocol consisted of 5 scans with a constant step size of 0.025\degree, decreasing \tth-range and increasing counting time per step as shown in Table \ref{tab:SCT-protocol}.

\begin{table}
    \caption{Staircase-counting-time (SCT) protocol for the lab PDF measurements on the Bruker D8 Discover diffractometer.}
    \begin{center}
    \begin{tabular}{lrrr}
    \toprule
         ~ & \tth start & \tth stop & counting time (sec) \\
    \midrule
         scan 1 & 2 & 140 & 1.8 \\
         scan 2 & 75 & 140 & 3.6 \\
         scan 3 & 107 & 140 & 7.2 \\
         scan 4 & 124 & 140 & 14.4 \\
         scan 5 & 132 & 140 & 28.8 \\
    \bottomrule
    \end{tabular}
    \label{tab:SCT-protocol}
    \end{center}
\end{table}

Synchrotron total scattering measurements were conducted at ID31 at the European Synchrotron Radiation Facility (ESRF) in Grenoble, France. 
The sample powders were loaded into cylindrical slots (1~mm thickness) held between Kapton windows in a high-throughput sample holder. 
Each sample was measured in transmission with an incident X-ray energy of 75.00~keV ($\lambda=0.1653$~\AA). 
A Pilatus CdTe~2M detector ($1679 \times 1475$ pixels, $172 \times 172$~$\mu$m each) was positioned with the incident beam in the corner of the detector. 
The sample-to-detector distance was approximately 0.3~m.
Background measurements for the empty windows were measured and subtracted. 
NIST SRM 660b (LaB6) was used for geometry calibration performed with the software pyFAI \cite{ashiotisFastAzimuthalIntegration2015} followed by image integration, including a flat-field, geometry, solid-angle, and polarization corrections. 
Data processing to obtain the PDF after any pre-processing to correct for absorption was done using \getx \cite{juhasPDFgetX3RapidHighly2013}. 
The Fourier transform from \fq to \gr was done with the parameters $\qmax=16.6$~\AA ,  $\qmin=1.0$~\AA, and \rpoly in the range 1.2-1.5. A \qmax of 30~\AA was used for the synchrotron data. 
The value of \rpoly was set according to normal protocols to ensure that the shape of the \fq function had a concave baseline whilst keeping \rpoly less than the nearest neighbor bond-length in the material.

Models were fit to the PDF data using \cmi \cite{juhasComplexModelingStrategy2015c}. 
As we did in our previous work \cite{schertenleibProtocolsObtainingReliable2024}, we fit models to uncorrected, corrected, and synchrotron data to evaluate the effect of the absorption correction by comparison to the fitted synchrotron data. 
Structural models for \ce{m-ZrO2}, \ce{c-CeO2}, and \ce{m-HfO2} were taken from the Inorganic Crystal Structure Database (ICSD) (entry IDs 80047, 184584, and 60902, respectively) \cite{bondarsPowderDiffractionInvestigations1995,scaviniProbingComplexDisorder2012,hannMonoclinicCrystalStructures1985}.
The fit was performed on the Nyquist-Shannon (NS) grid to facilitate propagation of valid estimated uncertainties\cite{tobyDeterminationStandardUncertainties2004}. 
The refined parameters include the scaling variable $s_1$, the damping factor \qdamp, the broadening parameter \qbroad, the correlated motion parameter $\delta_2$, the lattice parameters, and the atomic displacement parameters (ADPs). 
The ADP constraints and fitted $r$ range vary slightly for different datasets and will be presented separately for each dataset.

\section{Results}
\subsection{Assessment of the absorption correction for different \mud's}

\begin{figure}
    \centering
    \includegraphics[width= 0.65\textwidth]{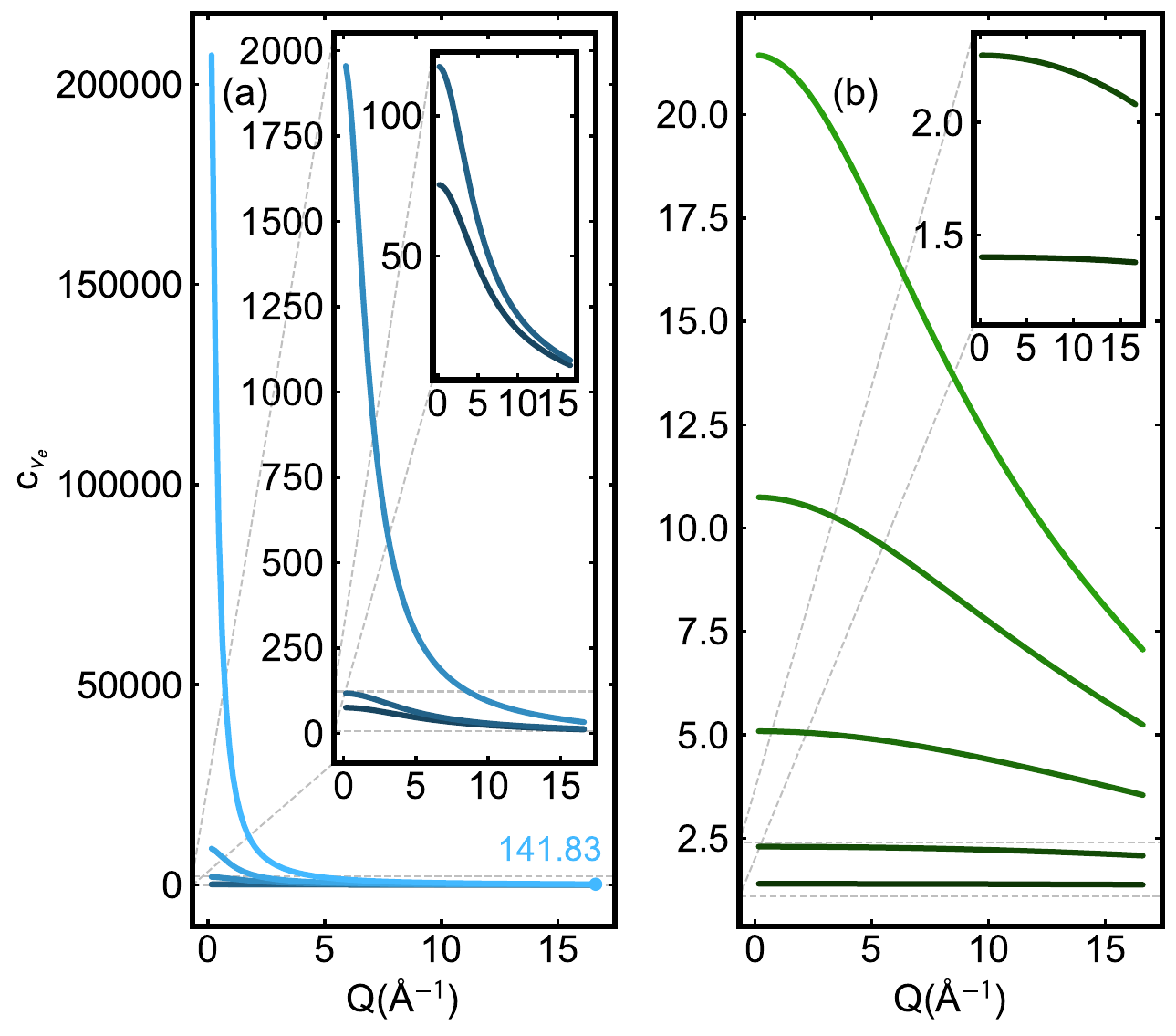}
    \caption{Absorption correction $c_{v_e}$ calculated using the brute-force method for various values of \mud, over the entire $Q$-range for Mo~\kalpha radiation. 
    From bottom to top in (a) $c_{v_e}$ for $\mud = 6.14$ (dark blue), 7.05, 17.15, 28.16, 71.51 (light blue), and (b) $c_{v_e}$ for $\mud =0.4$ (dark green), 1, 2, 3, and 4 (light green). The insets show the \cve curves for lower \mud's on expanded scales.
    In (a), the \cve value for the largest $Q$ at the highest \mud is labeled in the same light blue color as the corresponding curve.
    }
    \label{fig:cve}
\end{figure}

In \fig{cve}, by way of example, we show curves of $c_{v_e}$ for various choices of \mud.  
In \fig{cve}(a) \mud varies from 6.14 (lower dark blue curve) to 71.51. 

The \mud values are representative of samples of \ch{TiO2}, \ch{ZrO2}, \ch{SnO2}, \ch{CeO2}, and \ch{HfO2}, respectively, measured with Mo~\kalpha radiation in a 1~mm tube, but are only chosen as representative of real materials from different rows in the period table for comparison.

All the curves fall from a high value at low-$Q$ to a smaller value at high-$Q$.  
For the heavier elements the low-$Q$ signal would be multiplied by up to 200,000 times, falling to around 140$\times$ at the highest-$Q$ of 16.5~\iaa .
Experiments are obviously  not tenable with a 10$^5$ attenuation in signal and for heavier elements at modest x-ray energies thinner samples are needed, as is widely known.  
We explore this in more quantitative detail below.

In the smallest inset of \fig{cve}(a), we show the curves for more experimentally reasonable, albeit somewhat high, \mud values  of around 6.  In \fig{cve}(b) we show curves for \mud less than 6 ($0.4 < \mud < 4$).  
Even for these experimentally more realistic cases, the value of $c_{v_e}$ has a strong \q-dependence. For the higher range of \mud, the value of $c_{v_e}$ can change by $\approx 50\%$ from the low-\q to the high-\q end.  
A \mud in the vicinity of unity is generally considered optimal (shown in the inset to \fig{cve}(b)). 
Below we explore data quality for higher \mud values for the cases where it is difficult to make samples for PDF that are sufficiently thin.

\subsection{Exploration of different \mud cases}

\begin{figure}
    \centering
    \includegraphics[width= 1\textwidth]{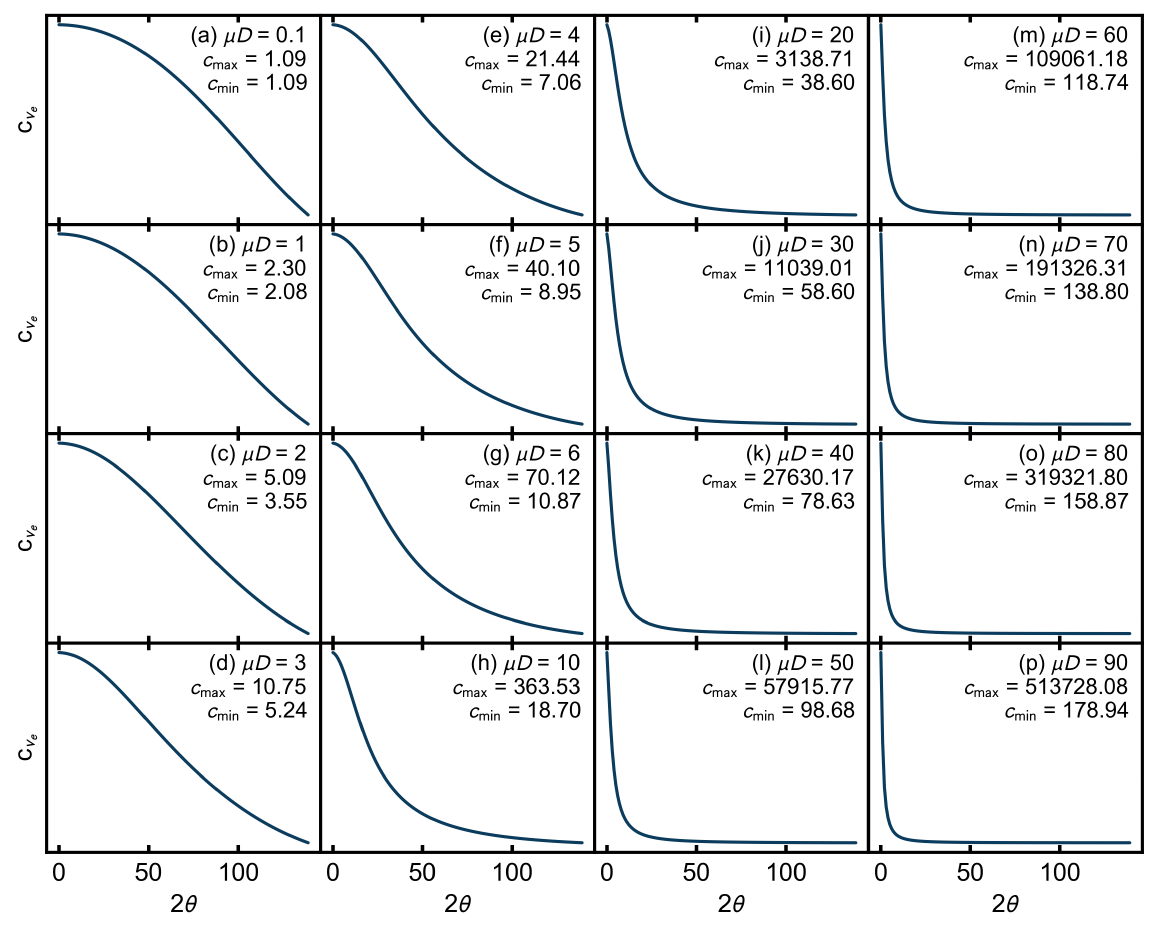}
    \caption{Comparison of the shape of absorption correction $c_{v_e}$ curves for the range of \mud values from 0.1 to 90. The $c_{v_e}$ curves are normalized from 0 to 1 to emphasize the shape of the curves vs $\tth$ for different values of \mud. The values of $c_{\max}$ and $c_{\min}$ reported in the panels are the maximal and minimal values, respectively, of $c_{v_e}$ before normalization for the angular range shown ($0<\tth<140$\degree). }
    \label{fig:all-cve}
\end{figure}

To understand how the $c_{v_e}$ curves vary in shape with \mud we compute them and plot them scaled to go from zero to one, in \fig{all-cve}.
They are normalized to go from zero to one as we are interested in the shape of the curve and not its absolute magnitude in this analysis.
The figure shows curves in the range from $0.1 < \mud < 90$. 
All the curves fall off more slowly, followed by a rapid fall-off with increasing \q and finally a long tail in the high-\q region.
What is characteristic of increasing \mud is that the crossover happens at lower angle, and therefore lower-\q. 
For the lower \mud values the curves are relatively flat over a much wider angular range.
The flatness of these curves for small \mud allows the absorption correction to be neglected in the \adhoc PDFgetX3 algorithm \cite{juhasPDFgetX3RapidHighly2013} for the case of high-energy synchrotron x-ray measurements in the RAPDF \cite{chupa;iucrcpd03} geometry.
In that geometry, the maximum \tth angle is around 40-50\degree, much less than the $\tth_{max} = 140 \degree$ in typical laboratory PDF measurements, which places the absorption correction even more in a flat region of the $c_{v_e}$ curves.
For lower-energy x-rays, which are more absorbing and require wider measurement angles, whether measured on laboratory instruments or at the synchrotron, this absorption correction should be considered in most cases.

The maximal and minimal absolute values of $c_{v_e}$ before the normalization are also reproduced in each panel of the figure to give an idea about how large of an angle dependence is required on an absolute scale.
For $\mu D \le 2$, even for the wide angular range of the data considered here, the angle dependence of the correction is quite small, but it grows rapidly for larger \mud values.
By a $\mu D = 6$ the low-\q correction is 7$\times$ larger than the high-\q correction and there is a significant angular dependence, though the data are probably still usable after applying the correction we lay out here.

To summarize, scattering properties are optimal and absorption corrections are minimal for \mud values of around 1.  
However, samples with \mud up to around 6 result in reasonable absorption corrections, but should have a correction applied for data collection over a wide angle.
For samples with larger \mud values the corrections become large and the data are not likely to be good.

\subsection{Estimating \mud for a sample}
\label{sec:diff_mud}

In this Section we consider a number of different ways for estimating \mud for a sample.
At first sight this is straightforward since this quantity can, in principle, be calculated from mostly known quantities.
However, in practice, some quantities such as the sample density or its chemical composition, are often not well known.  
Also, the model we used to compute the correction makes some assumptions that are not necessarily true in practice. For example, that of a parallel beam of the same width as the sample is not necessarily true in practice.  
We therefore seek to determine an ``effective" \mud, \mude, for our sample which is the \mud that gives the most appropriate \cve curve given our experimental conditions.
We compare a number of different approaches for determining \mude by evaluating their effect on the refined structural parameters.
We would like to understand empirically which approaches for estimating \mud are preferred, as well as, in general, how big of an effect the \cve correction has on refined parameters.

\begin{figure}
    \centering
    \includegraphics[width=\linewidth]{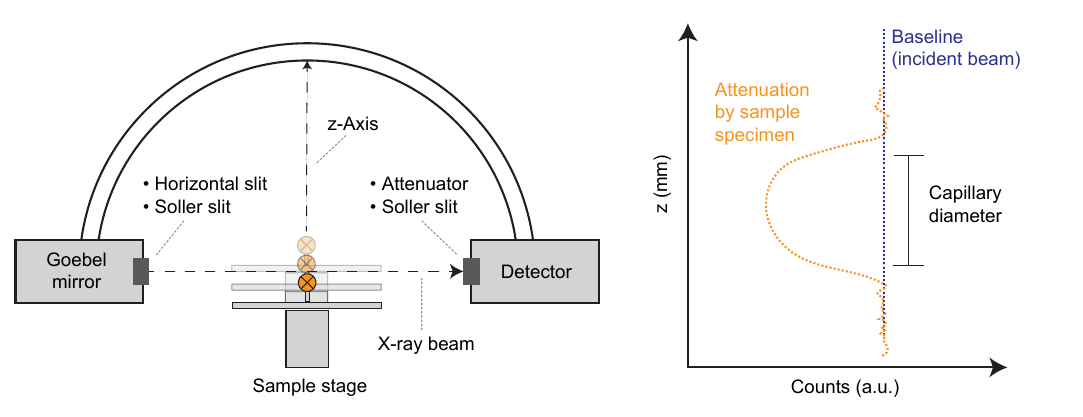}
    \caption{Scheme of the diffractometer setup for direct measurements of the sample \mud, called here \zscans. 
    The sample stage is moved along $z$ so the sample traverses the  incident beam whilst recording the  x-ray intensity as a function of $z$. 
    The capillary sample is displayed as an orange circle, with its principle axis perpendicular to the viewing plane. 
    This results in a U-shaped profile (orange curve on the right). 
    The capillary diameter, marked by the black bar on the right, is slightly shorter than the full opening of the orange U-shaped curve because of the finite width of the beam.}
    \label{fig:zscan_scheme}
\end{figure}

We first consider a number of different ways for estimating \mud for a sample. Later we compare them by modeling.
The \mud of a sample can be measured directly for a given diffractometer setup by measuring the x-ray attenuation of the capillary specimens placed in the incident beam as shown in \fig{zscan_scheme}. 
In this measurement, the sample stage is moved vertically (defined here as the $z$-axis) traversing the incident beam. 
In our case, for this, the Bruker Lynxeye detector is set to 0D-mode, which means that the whole detected area is integrated.

The resulting absorption profile can be fit with a function that is a convolution of the width of the incident beam and the absorption characteristics of the sample, assuming it is cylindrical, as we describe below.  

We tested absorption scans where the incident beam height was made narrower and more tightly collimated than was the case for the actual measurement of the PDF data.
We also tested making the absorption scan with wider incident beam heights, including the height used in the actual PDF measurements.
To accomplish the experiment with the narrower beam, either the active channels on the detector were reduced, or a horizontal slit  was placed in the incident beam before the sample to narrow the beam. 
The former limits the number of pixels that are active, limiting the active area on the detector to a horizontal stripe. 
The latter masks the beam and limits beam divergence. 
We found that both have a similar effect on the resulting U-shaped \zscan. We then compared three methods to determine \mud from the \zscans.

\textit{Method 1:} An approximate \mud is determined using 
\begin{equation}
\mudmone = \ln(I_{\max}/I_{\min}),
\end{equation}
where $I_{\max} \approx I_0$ and $I_{\min} \approx I_0 \cdot e^{-\mud}$.
The rationale is that the minimum attenuation ($I_{\min}$) is reached when the center of the capillary aligns with the center of the x-ray beam. 
If the vertical beam height is small enough relative to the capillary diameter, then at this point the sample thickness doesn't change much across the beam diameter and the measurement resembles a standard $\mu t$ measurement of a sample of uniform thickness $t$.  Knowing $\mu t$ at the position of the diameter, and the diameter, we can obtain $\mu$, and therefore \mud for the capillary.
This reasoning breaks down when the beam height becomes comparable to the sample height.

\textit{Method 2}: We carry out a fit of the \zscan curve to a model that assumes a circular cross-section capillary of uniform density and a parallel x-ray beam of height $h$, which estimates \mud. The mathematical details are included in Supplementary Section~\ref{sec:fit_mud_si}. 
The fit yields \mudmtwo by fitting parameters $\mu$, $D$, $h$, $I_0$, $z_0$, the height of the center of the capillary, and $m$, a linear coefficient for $I_0$ that we found depends on $z$.
This model ignores effects of the sample container which we expect to have a negligible effect for thin polyimide or quartz tubes at Mo or Ag~\kalpha energies.  
It will be a less good approximation if the sample container absorption is significant.

\textit{Method 3:} Method 3 is the same as Method 2 except that for \mudmthree the capillary diameter, $D$, is fixed to the known value from the manufacturer and not allowed to vary as it is in Method 2. 
In an ideal world, Method 2 would return a fit diameter that is very close to the known physical diameter, but we found that this was not always the case due to inadequacies 
in our model and we wanted to understand the effect this has on the results. 

It is also possible to estimate \mud theoretically using the known x-ray wavelength, sample composition, and densities and/or powder packing fractions. and so we define also \textit{Method 4:} This makes use of tabulated attenuation coefficients from online resources such as XrayDB \citexraydb, APS \cite{vondreeleComputeXrayAbsorption2024}, and the NIST Standard Reference Database 126 \cite{hubbellTablesXRayMass2004}. 
Here we calculate \mudmfour values using a measured mass density for our loaded samples and the lookup tables that makes use of the python XrayDB database \citexraydb. The computing details are described in Supplementary Section~\ref{sec:theoretical_mud_si}. 

\begin{figure}
    \centering
    \includegraphics[width= \textwidth]{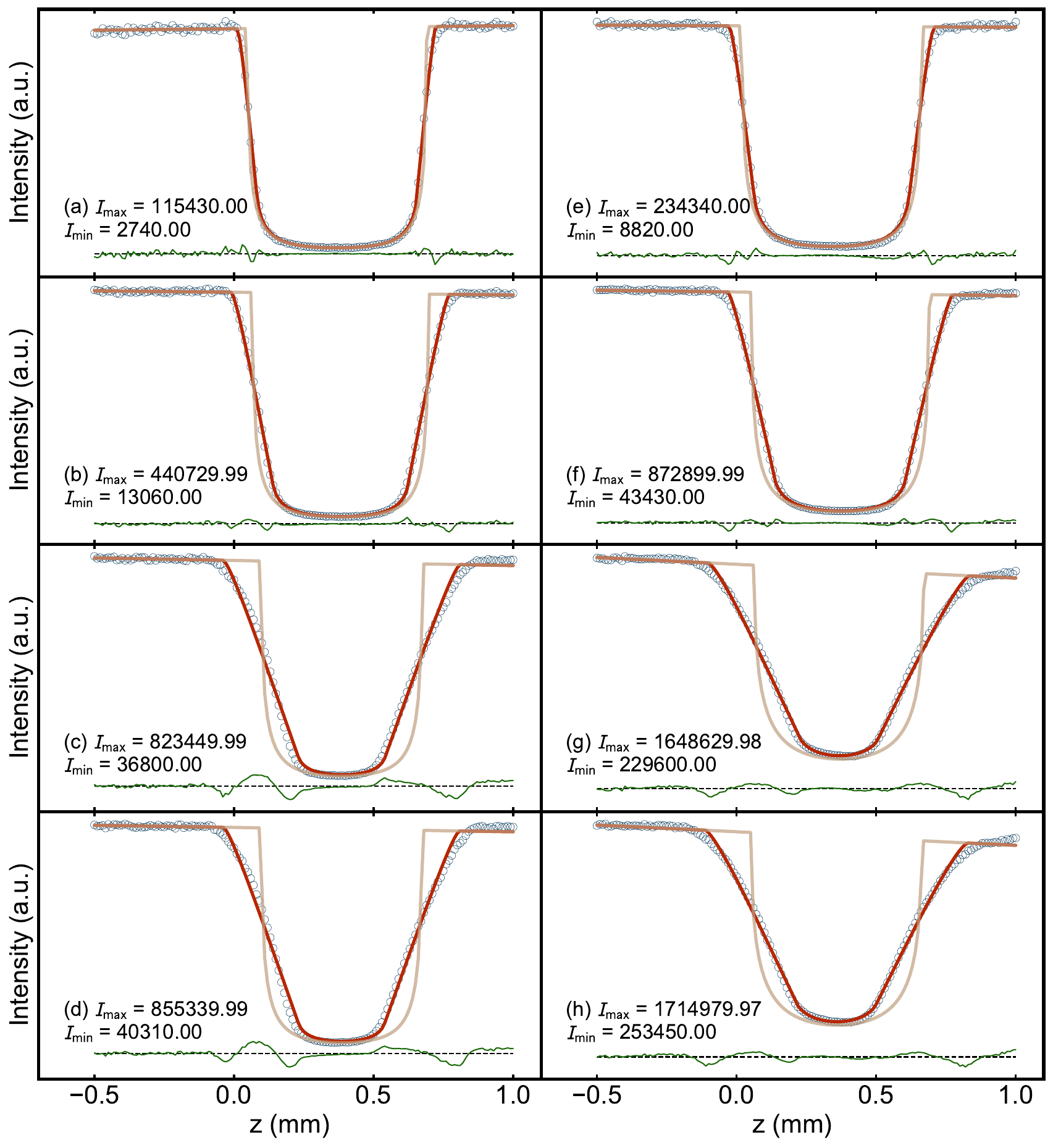}
    \caption{Examples of \zscans computed using Method 2 (blue circles) with model fits (red lines) for \ce{CeO2} with ID $= 0.635$~mm. The green lines offset below show the residuals. In each case the brown U-shaped curve is the unconvoluted intensity which gives an indication of the edges of the capillary.
    The panels are arranged so side-by-side panels are measured with closed (left) and open (right) detector channel settings, and measured with increasing height of beam slit going down the columns $h=0.05$~mm, 0.2~mm, 0.6~mm, and 1~mm.}
    \label{fig:CeO2_zscan_fit}
\end{figure}

For convenience we have developed \mud calculators in the Python software package \utils \cite{farrowDiffpyutils2024} for Methods 2 and 4, which are free to use.
For Methods 2 and 3, we also explored various experimental settings for the \zscan, including varying x-ray beam heights $h$ and detector channel conditions.
The full set of \mud values and fitted parameters that we tried is presented in Tables~\ref{tab:mud_list} and \ref{tab:fitted_parameters} in Supplementary Information. Here we confine ourselves to \zscan data for \ce{CeO2} with ID = 0.635~mm which was chosen since it had a high, but not excessive, \mud (\mudmfour = 4.229), which, according to the \mud curves in \fig{all-cve}, suggests that an inappropriate estimation of this will result in a significant effect on the measured intensities. Examples of measured \zscans are shown in \fig{CeO2_zscan_fit} as blue circles, computed using Method 2, with fits shown as red lines.
The theoretical curve for the sample transmission before convolution with the beam height is shown as the brown curves.

When comparing across rows where $h$ is constant, we observe that the maximal intensity $I_{\max}$ is around twice as large for open channels, while the minimal intensity $I_{\min}$ is about 4 times larger. 
This results in lower \mudmtwo's for open detector channels.
In addition, as $h$ increases (comparing down columns), we find that the slope of the U-shape curve (blue curve) becomes less steep. 
In all cases the fits of our model are satisfactory, albeit returning quite different  \mudmtwo's.

\begin{figure}
    \centering
    \includegraphics[width= 0.8\textwidth]{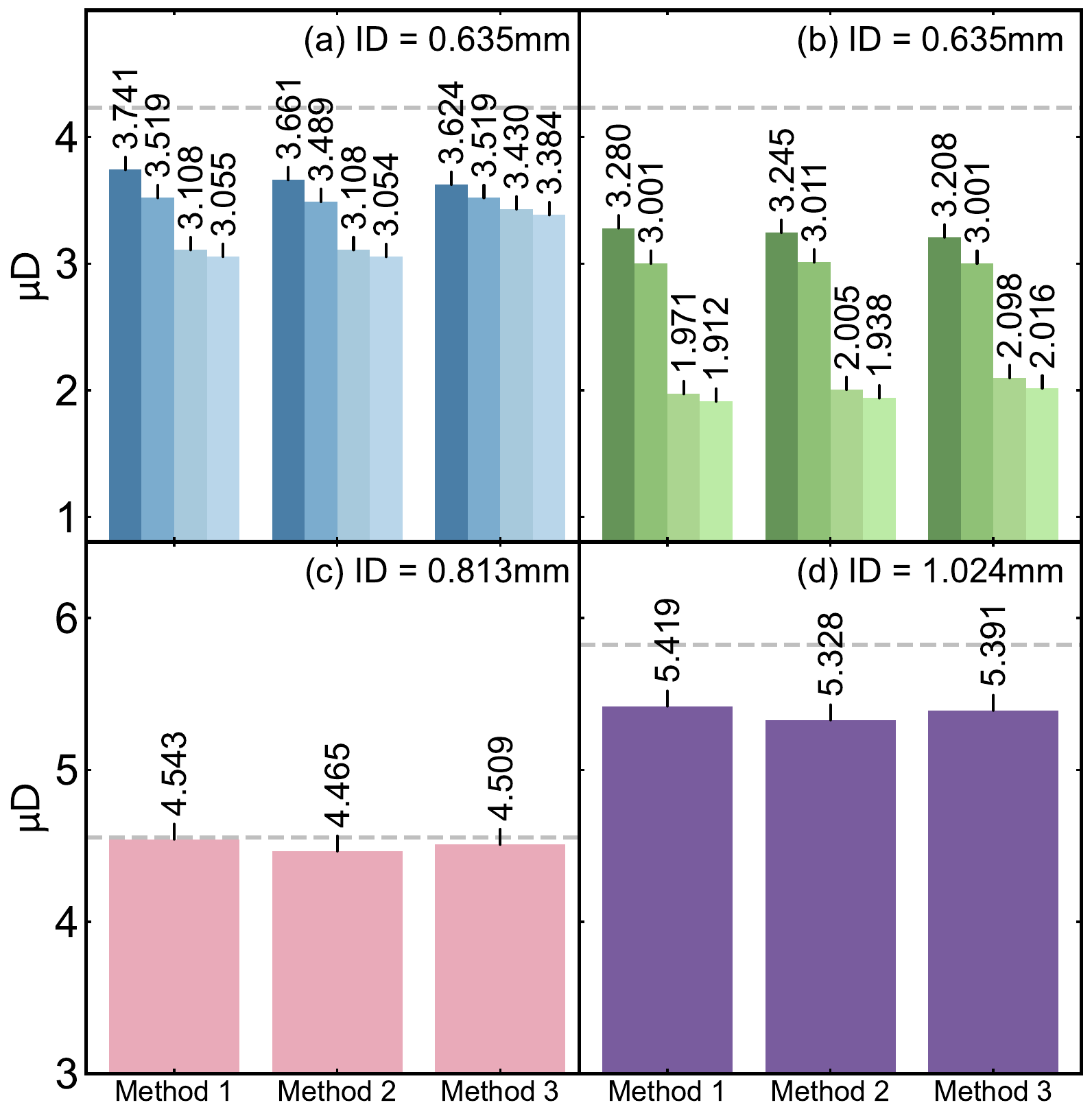}
    \caption{
    \mud values for different \ce{CeO2} datasets determined from \zscans using varying incident beam heights ($h$) and channel conditions. 
    In all panels, the IDs are reported, with \mudmone, \mudmtwo, \mudmthree, and \mudmfour shown.
    \mudmfour for panels (a), (b), (c), and (d) are 4.229, 4.229, 4.554, and 5.824, respectively, as indicated by the dashed grey line.
    In panels (a) and (b), lighter color bars indicate larger~$h$ values, $h=0.05$~mm, 0.2~mm, 0.6~mm, and 1~mm. Panel (a) shows the reduced channel detector condition, while (b) shows the open channel detector condition.
    Panels (c) and (d) show data with reduced channels with $h=0.1$~mm.
    }
    \label{fig:CeO2_barplot}
\end{figure}

The different \mudmtwo's are summarized in \fig{CeO2_barplot} (a) and (b), along with those of \mudmone and \mudmthree, for the same experimental conditions considered in \fig{CeO2_zscan_fit}.

Comparing across the methods, we see that for each experimental setting indicated by the same color, Methods 1-3 give very similar values as each other.
This means that we could use any of these methods to estimate \mud and it wouldn't matter.
On the other hand, Method 4, using the theoretically determined \mud, indicated by the dashed grey line, gives a consistently higher value.

However, we see large variations in the estimate of \mud depending on the experimental conditions used to measure it.  Comparing across color lightness (beam height, $h$) and between panels (a) and (b) (closed vs. open detector channels), the resulting \mud can vary by almost a factor of two.

A question therefore arises, which of these values of \mud is the most appropriate for correcting the XRD data for PDF analysis, and that is explored in the next Section.
Here we can make some observations that help us to understand why we see these variations.

We first note that the beam height, limited by physical slits upstream of the sample and the channel settings of the detector have a similar effect.  
The former limits the effective area of the sample that is illuminated, and the latter limits the area of the beam on the detector that is accounted for.
Comparing panels (a) and (b), we always get slightly lower \mud values with the open-channel setting, but the difference is small when $h$ is small (i.e., 0.05 and 0.2~mm).

We note that another important length-scale in the problem which is the diameter of the capillary (0.635~mm in this case).
When $h$ is much smaller than the capillary diameter, we get larger estimates of \mud that are also closer to the theoretical \mud's but much smaller estimates of \mud when the $z$-scan is measured with beam-slit heights comparable to or larger than the sample capillary. 

\subsection{Understanding the effect of different \mud correction magnitudes on fits and refined parameters}

We next investigate the effect on the fit-quality and refined parameters of correcting the ceria PDFs using various different \mud's.
We use \ce{CeO2} data from the $1.024$~mm capillary as an example, where the absorption effect is strong. 
For the purpose of comparison, in addition to the uncorrected data, we consider three different \mud values for the corrected data: the experimental $\mud = 5.328$ obtained from the \zscan with $h=0.1$~mm and reduced channels using Method 2, a purposefully underestimated \mud = 3, and an overestimated \mud = 10.  
We do this by computing structure functions and PDFs from the measured data after each of the three corrections, and fitting the PDFs of the ceria crystal structure model using \cmi.

\begin{figure}
    \centering
    \includegraphics[width=\textwidth]{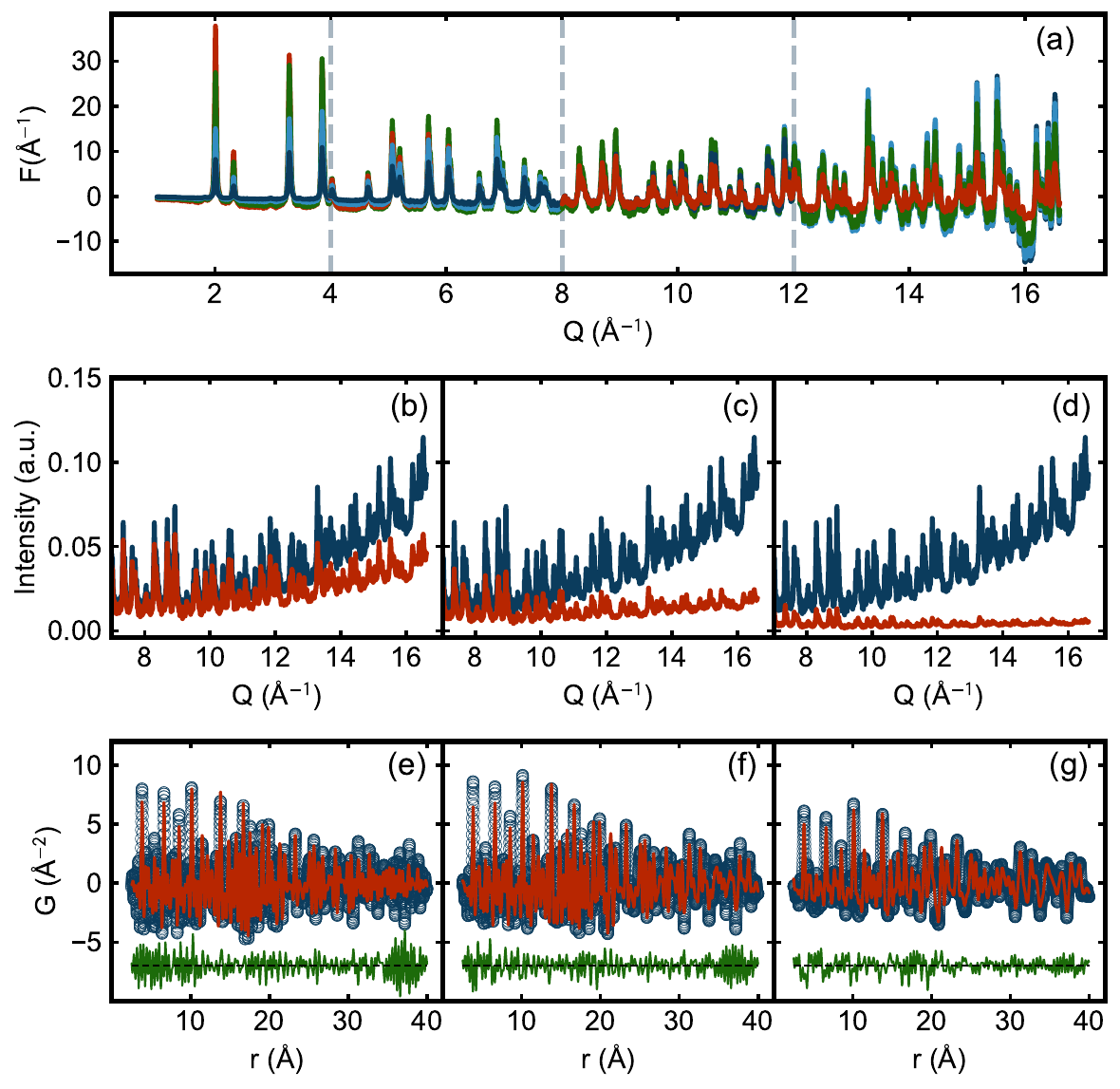}
    \caption{
    Comparison of \ce{CeO2} data (ID = 1.024~mm) with and without absorption correction, using small (\mud = 3), middle (\mud = 5.328), and overestimated (\mud = 10) values. 
    Panel (a) shows \fq, where the dark blue, light blue, green, and red curves represent no correction and corrections with small, middle, and overestimated \mud, respectively. 
    The curves overlap or change in amplitude across \q-ranges, marked by the light grey dashed vertical line, with the order of plotting adjusted to reflect changes without disrupting overall trends.
    Panels (b)-(d) show the original intensity (blue) vs. corrected intensity (red), scaled so that the highest intensity peak equals to one. Plots are shown on a \q scale, focusing on $Q > 7$~\iaa.
    Panels (e)-(g) show the best-fit calculated PDFs (red lines) on top of their respective measured, absorption corrected, PDFs (blue circles).
    The PDFs were fitted between $r_{\min} = 1.0$ and $r_{\max} = 40.0$. 
    The corrections in left columns (b) and (e) correspond to the small \mud, the middle columns (c) and (f) to the middle \mud, and the right columns (d) and (g) to the overestimated \mud.}
    \label{fig:CeO2_correction}
\end{figure}

The effect of these corrections on the raw data, \fq, and \gr are shown in 
\fig{CeO2_correction}.

We consider first the effect on the high-\q region of the raw data (\fig{CeO2_correction}(b-d)).
In these plots the curves are normalized so the strongest peak in each curve  is set to unity, allowing us to conveniently compare the behavior at high-\q.
Similar to what was shown in Schertenlieb \textit{et al.}~\cite{schertenleibProtocolsObtainingReliable2024}, albeit in that case for lower absorbing samples, we see that the absorption correction reduces an overall upturn in the data whilst at the same time suppressing the high-\q signal.
In other words, neglecting the correction will result in an amplified signal at high-\q, which might resemble an under-estimated Debye-Waller factor in subsequent refinements.
Unsurprisingly, the effect on the signal becomes larger as the \mud correction is increased, though the slight upturn in the data doesn't completely go away even for $\mud = 10$ which we believe to be a significant overestimate of the actual sample absorption. 
As discussed briefly before~\cite{schertenleibProtocolsObtainingReliable2024}, there is an additional contribution to the upturn that is not coming from absorption effects and so simply removing the upturn cannot be used to estimate \mud.

The effect on \fq plotted over the whole range of~\q is shown in \fig{CeO2_correction}(a), where we find under-estimation in low-\q signal and over-estimation in the high-\q signal when comparing uncorrected to corrected data. 
The progressive suppression of the high-\q signal (\q$>12$) with increasing correction is very clear.

We turn to fits of structural models to the PDF to understand the effect of the corrections on the fit quality and on refined structural parameters. 

The \gr fits for the corrected data are shown in \fig{CeO2_correction} (e)-(g). A larger version is shown in \fig{ceo2_pdfs} in the Supplementary Information.
All the curves result in reasonable model fits, with small misfits, as evident in the green difference curves.
The \rw value for no correction is approximately 0.72, and it improves to 0.6, 0.44, and 0.36 for small, middle, and overestimated \mud, respectively. 
The fit quality, as measured by the \rw, and visually in the difference curve, gets better with increasing magnitude of the absorption correction.
This trend occurs even when we use a correction that is known to be too strong.
This means that fit quality alone, as measured by \rw, is not a good measure of the best absorption correction to use.
Of course, the purpose of the fits is to obtain high quality values for refined structural parameters.
We therefore turn to an analysis of the refined parameter values obtained with different degrees of absorption correction.

The refinement results for the three selected \mud corrections are shown in Table~\ref{table:CeO2_diff_muD_fits}. 
We are most concerned about the refined values of structural parameters that are, in this case, $a$, Ce(\uiso) and O(\uiso). 
The effect of the correction on the lattice parameter is small, varying at the thousandths of an \AA nsgr\"{o}m level, but does show a small decrease in lattice parameter with increasing strength of correction.
The  Ce(\uiso) and O(\uiso) values vary monotonically with \mud.
Ce(\uiso) increases with increasing absorption correction in line with the progressive suppression of high-\q signal by the correction.  
Surprisingly, O(\uiso) shows the opposite behavior, decreasing with increasing correction magnitude, but this may reflect the fact that the oxygen contributions to the measured PDFs in \ch{CeO2} is very small and so these ADPs are not very reliable.

To decide which \mud values are optimal we compare to fits to the synchrotron data.
The refined lattice parameters from the lab data are closer to reported values\cite{scaviniProbingComplexDisorder2012}, indicating that the synchrotron data overestimates the lattice parameter, which may be related to a calibration issue with the synchrotron data. 
Among the three selected correction strengths, we find that the experimentally determined $\mud = 5.328$ provides \uiso values that agree the best with the synchrotron data.

\begin{table}
\begin{center}
\caption{Results of the refinement for \ce{CeO2} data (ID=1.024~mm) for synchrotron, uncorrected, and corrected data with selected \mud's, over $r_{\min} = 1.0$ and $r_{\max} = 40.0$. The \mud's are the values used for each correction.}
\label{table:CeO2_diff_muD_fits}
\begin{tabular}{llllll}
\toprule
Parameter & synchrotron & uncorrected & \mud = 3 & \mud = 5.328 & \mud = 10 \\
\midrule
       $s_1$ &                       0.37434(21) &                      0.2191(9) &                                     0.2972(12) &                                         0.3693(12) &                                       0.3287(9) \\
    \qdamp &                        0.02386(4) &                    0.03186(25) &                                    0.03074(24) &                                        0.02737(23) &                                     0.01470(32) \\
   \qbroad &                        0.01814(7) &                      0.0220(5) &                                      0.0251(5) &                                          0.0320(4) &                                       0.0461(4) \\
   $\delta_2$ &                           9.07(4) &                       9.92(28) &                                       9.21(17) &                                           7.87(13) &                                        4.99(13) \\
        $a$ &                       5.414232(7) &                     5.40240(5) &                                     5.40160(5) &                                         5.40018(5) &                                      5.39814(6) \\
   Ce(\uiso) &                       0.003571(5) &                   0.002403(24) &                                   0.002671(25) &                                       0.003154(29) &                                      0.00467(4) \\
   O(\uiso) &                       0.04120(10) &                     0.0990(24) &                                     0.0696(13) &                                          0.0472(7) &                                       0.0353(4) \\
     $Q_{\max}$ &                              30.0 &                           16.6 &                                           16.6 &                                               16.6 &                                            16.6 \\
     grid &                           0.10472 &                       0.189253 &                                       0.189253 &                                           0.189253 &                                        0.189253 \\
       \rw &                          0.160823 &                       0.718991 &                                       0.596404 &                                           0.439131 &                                        0.363539 \\
  $\chi_{red}^2$ &                        961.528958 &                    1173.593401 &                                     753.497185 &                                         447.645936 &                                      452.603539 \\
\bottomrule
\end{tabular}
\end{center}
\end{table}

A detailed exploration of \mude for all \ce{CeO2} data is presented in Supplementary Section~\ref{sec:select_mude}, where we propose a recommended reference value. 
Additional examples for both low and high sample absorption cases are discussed in Supplementary Section~\ref{sec:low_and_high_mud}.

\section{\labpdfproc software for absorption corrections}

The corrections described here have been implemented in an easy to use Python package, \labpdfproc, as part of the \diffpy\cite{DiffPyAtomicStructure} family of software packages.
It can be used to compute absorption corrections and apply them to measured data before the data are fed to, for example, \getx, to obtain PDFs. 

The software can be easily installed from the Python Package Index (Pypi)\cite{PythonPackageIndex} or from conda-forge~\cite{Condaforge} and so is straightforward to use alongside other \diffpy programs. 
Full instructions for installation and use can be found at the GitHub readme\cite{Diffpylabpdfproc2024} and the online documentation\cite{DiffpylabpdfprocDocumentation2024}.

To use this tool, you only need your 1D~diffraction pattern data, along with either the \mud value, a \zscan file, or relevant chemical information. 
The program is fast and easy to use.
By default, we use a fast calculation for \cve correction, described in Supplementary Section \ref{sec:fast_calc}.  
It can be run on single file or on directories of files.  
It also implements the more computationally intensive brute-force calculation method that is described in Section~\ref{sec:abs_corr}.
\labpdfproc can also be used to facilitate estimating \mud for a sample of known composition before making the samples as part of the experiment design.

\section{Recommended Protocols for sample preparation and absorption correction for PDF measurements in laboratory settings}

Overall, we find that making an absorption correction is better than not, especially for $2 < \mud < 7$. For high $\mud > 8$, we are unlikely to get usable data. 
Here, we provide our recommended protocols for handling sample absorption in detail. 
For other experimental decisions for PDF experiments from lab sources, please refer to our previous work \cite{schertenleibProtocolsObtainingReliable2024}.
The linear absorption coefficient, $\mu$, can be estimated if an approximate composition is known by using our program or using websites mentioned herein. If you do not know the packing density of your sample, use an approximate value of 0.5. 
Choose a sample diameter to give a \mud of less than 2 ideally but anything up to 6 or even 7 is acceptable.
For very absorbing samples, where no tube or capillary is available that is thin enough, you can consider coating the outside of a narrow tube or glass fiber in grease and sprinkling a thin layer of powder on the outside.
The results of doing this for our \ch{HfO2} sample are reproduced in the Supplementary Information Section~\ref{sec:low_and_high_mud}.

To estimate the most appropriate \mud for correction, \mude, we propose the following:
(1) If the capillary diameter is close to the x-ray beam height ($D \geq 0.8 * h$), either compute \mud theoretically or measure the experimental \mud using a \zscan with reduced channels and small beam heights. 
(2) If the capillary diameter is much smaller than the beam height, scale the theoretical \mud with $D/h$ or do a \zscan with the actual experimental settings.

\section{Conclusion}
In contrast to rapid acquisition PDF experiments~\cite{chupa;iucrcpd03}, PDF experiments carried out on laboratory diffractometers, or at synchrotrons but with low x-ray energy, require absorption corrections.
Here we revisit the nature of absorption corrections, and their effect on the resulting structural refinements, for laboratory PDF measurements which are becoming more popular.
We report a new software program, \labpdfproc, that can carry out absorption corrections in a fast and straightforward way before data are then propagated to the popular \getx program.
We also assessed the effects of different absorption corrections on subsequent PDF refinements, as well as exploring different methods for determining the best \mud values in a particular experimental situation.
The goal is to make this step as straightforward as possible so that it can be easily incorporated into laboratory PDF experimental workflows.
As a result, we present protocols for sample preparation to improve data quality.

\section{Acknowledgements}

Synchrotron total scattering measurements were performed at beamline ID31 at the European Synchrotron Radiation Facility (ESRF).
Work in the Billinge Group was supported by U.S. Department of Energy, Office of Science, Office of Basic Energy Sciences (DOE-BES) under contract No. DE-SC0024141. T. Schertenleib was supported by the Swiss National Science Foundation under grant number 200021\_188536.

\section{Data availability}

Data used for all the plots in the manuscript are available on Zenodo at \url{https://doi.org/10.5281/zenodo.15199572}.

\bibliography{bg-pdf-standards,billinge-group-bib,yc_absorption_correction,ts_mo_data_reduction,yc_absorption_correction_web_citations}

\pagebreak
\begin{center}
\textbf{\Large Supplementary Information}
\end{center}
\setcounter{equation}{0}
\setcounter{figure}{0}
\setcounter{table}{0}
\setcounter{section}{0}
\makeatletter
\renewcommand{\thefigure}{S\arabic{figure}}
\renewcommand{\thetable}{S\arabic{table}}
\setcounter{figure}{0}
\makeatletter 
\let\c@table\c@figure
\makeatother
\renewcommand*{\thefigure}{S\arabic{figure}}
\renewcommand*{\thetable}{S\arabic{table}}

\section{Derivation of the path length of an x-ray through the sample}
\label{sec:path_length_si}

We derive here the relations for determining $\ell_v(\tth)$, the path length through the sample of an x-ray scattered at angle \tth from the $v$th voxel.
As shown in \eq{ell_tth}, this is given by the sum of the path lengths of the incoming and outgoing x-ray paths, $\ell_v^i$ and $\ell_v^o$, respectively. 

In our simplified model we assume that the incoming beam is parallel (non-diverging).
The geometry is shown schematically in  \fig{path_lengths} of the main paper, where the circle in this figure represents the cross-section of the capillary and the incoming beam arrives horizontally from the left.
Placing the origin of our coordinate system at the center of the circle we can consider a pixel, $v$, located at position  with coordinates ($x_v$,$y_v$) (as examples, see the large grey dots in \fig{path_lengths}).
The incident path length will be given by $\ell_v^i = |x_v - x_i|$, where the coordinates where the beam enters the capillary circle are ($x_i$, $y_i)$ (indicated by the red dots in \fig{path_lengths}).
Points on the capillary circle satisfy $x^2 + y^2 = R^2$, where $R = \frac{D}{2}$ is the radius of the capillary circle.
Since $y_i = y_v$ which is known, we take the negative value and get that $x_i = -\sqrt{R^2-y_v^2}$ and
\begin{equation}
    \ell_v^i = \left|x_v + \sqrt{R^2-y_v^2}\right|.
    \label{eq:ell_in}
\end{equation}

To get the path length of the out-going x-ray we notice that, if the coordinates of the point on the circle where the ray leaves the capillary are ($x_o$,$y_o$) (indicated by the red dots in \fig{path_lengths}), then 
\begin{equation}
{\ell_v^o(\tth)} = \sqrt{(x_o - x_v)^2+(y_o - y_v)^2}.\label{eq:ell_out}
\end{equation}
Since $x_v$ and $y_v$ are known, it remains to find $x_o$ and $y_o$.

When $\tth \neq 90\degree$, we find the yellow line equation $y=ax+b$ that passes through both the voxel and the exit point, where its slope $a=\tan(\tth)$ is determined by the scattering angle and its intersection point can be computed from the grid points $b=y_v-x_v \cdot \tan(\tth)$. Therefore, we have that $x_o = \frac{y_o-b}{a}$. Since the exit point is an intersection point that passes through the line and the circle, we solve by establishing simultaneous equations for the circle and the line. We then have:
\begin{align}
    x_o^2 + y_o^2 &= R^2 \\
    \left(\frac{y_o-b}{a}\right)^2 + y_o^2 &= R^2 \\
    (1+a^2)y_o^2 - 2by_o + (b^2-a^2R^2) &= 0.
\end{align}

Since $a^2 \geq 0$, $1 + a^2 \neq 0$, and we can use the quadratic formula to get $y_o$:
\begin{align}
    y_o &= \frac{2b \pm \sqrt{4b^2-4(1+a^2)(b^2-a^2R^2)}}{2(1+a^2)} \\
    &= \frac{b \pm a \sqrt{(1+a^2)R^2-b^2}}{1+a^2}.
\end{align}

As we consider only voxels within the cylinder ($x_v^2 + y_v^2 \leq R^2$), the discriminant of this quadratic is positive:
\begin{align}
    (1+a^2)R^2 - b^2 &= R^2 + a^2R^2 - (y_v - ax_v)^2\\
    &= (R^2 - y_v^2) + (a^2R^2 - a^2x_v^2) + 2ax_vy_v\\
    &\geq x_v^2 + a^2y_v^2 + 2ax_vy_v\\
    &= (x_v + ay_v)^2\\
    &\geq 0.
\end{align}

Since we only have to consider paths leaving the circle in direction of the detector, the exit point is above the grid point thus is the upper intersection point. So we take the positive value here.
Substituting and rearranging the equations, we get:
\begin{align}
    x_o &= \frac{ \left(\sqrt{(1+\tan^2(\tth))R^2-(y_v-x_v\tan(\tth))^2} -y_v \tan(\tth) +x_v \tan^2(\tth) \right)}{1+\tan^2(\tth)} \\
    y_o &= \frac{\left(y_v -x_v \tan(\tth)
    + \tan(\tth)\sqrt{(1+\tan^2(\tth))R^2-(y_v-x_v\tan(\tth))^2} \right)}{1+\tan^2(\tth)},
\end{align}
which can then be plugged back into \eq{ell_out}.

This function is not well behaved in the vicinity of $\tth = 90\degree$.  At $\tth = 90\degree$, $x_o=x_v$ and so $y_o = \sqrt{R^2-x_v^2}$, and
\begin{equation}
    \ell_v^o (\tth=90\degree) = \left|y_v - \sqrt{R^2-x_v^2}\right|.
    \label{eq:ell_out_90_deg}
\end{equation}

\section{Derivation of the result that \cve depends on the product \mud: Express $\ell_v(2\theta)$ in radius $R$}\label{sec:mud_si}

From our previous expressions for $\ell_v(\tth)$, \eq{ell_in}, \eq{ell_out}, \eq{ell_out_90_deg}, we find that if we can express $x_v$ and $y_v$ linearly in terms of $R$, then we can factor $R$ out for all $\ell_v(\tth)$ and therefore use the product \mud as an independent variable of \cve rather than $\mu$ and $D$ separately.

Here, we present a method for expressing the Cartesian coordinates $(x_V,y_V)$ of any point $V$ inside a circle in terms of the radius $R$. This process is also visualized in \fig{varying-diameter} below. Consider the line segment $OV$ from the origin, $O$, at the center of the circle to the point $V$ inside the circle.  Draw a line perpendicular to $OV$ that goes through $V$.  This line intersects the boundary of the circle at two points. Choosing one such point and denoting it as $A$, we establish the radius $OA=R$. Let $\alpha_V = \angle VOA$, then $\cos(\alpha_V) = OV / OA$, which gives us $OV = R \cos(\alpha_V)$. Transitioning to polar coordinates, we have that $x_V = OV \cos(\beta_V)$ and $y_V = OV\sin(\beta_V)$, where $\beta_V$ represents the angle between the positive $x-$axis and $OV$. We therefore have $x_V = R \cos(\alpha_V) \cos(\beta_V)$ and $y_V = R \cos(\alpha_V) \sin(\beta_V)$.

\begin{figure}
    \centering
    \includegraphics[width= 0.5 \textwidth]{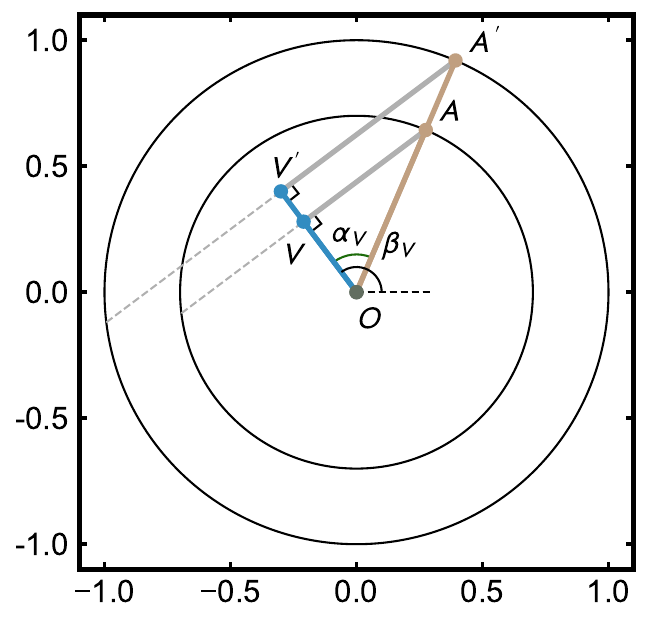}
    \caption{Visualization of the same $V$'th pixel for circles with varying diameters. The pixels are denoted by the blue dots, $V$ and $V'$, for the small and large circle, respectively. The blue lines represent the line segments connecting each pixel to the origin, and the dashed grey lines are the perpendicular lines intersecting each circle at two points. Here, we pick the light brown upper intersection point as $A$ and $A'$. These light brown lines thus represent the radii. We can then find $\alpha_V$ and $\beta_V$, which are the same for $V$ and $V'$.}
    \label{fig:varying-diameter}
\end{figure}

Denote $\alpha_V$ as the angle formed between the line segment connecting the $V$'th pixel to the origin and the radius of the circle to the origin, and $\beta_V$ as the angle between the positive $x$-axis and the line segment connecting the pixel to the origin.
Since $x_V = R \cos(\alpha_V) \cos(\beta_V)$ and $y_V = R \cos(\alpha_V) \sin(\beta_V)$, we have the following:
\begin{align}
    \ell_V^i &= \left|R \cos(\alpha_V) \cos(\beta_V) + \sqrt{R^2-(R \cos(\alpha_V) \sin(\beta_V))^2}\right| \\
    &= \left|\cos(\alpha_V) \cos(\beta_V) + \sqrt{1- \cos^2(\alpha_V) \sin^2(\beta_V)}\right| R,
\end{align}
\begin{align}
    \ell_V^o (\tth=90\degree) &= \left|R \cos(\alpha_V) \sin(\beta_V) - \sqrt{R^2-(R \cos(\alpha_V) \cos(\beta_V))^2}\right| \\
    &= \left|\cos(\alpha_V) \sin(\beta_V) - \sqrt{1-\cos^2(\alpha_V) \cos^2(\beta_V)}\right| R,
\end{align}
and
\begin{align}
    \ell_V^o(\tth \neq 90\degree) &= \sqrt{(x_o - x_V)^2+(y_o - y_V)^2} \\
    &= \sqrt{G^2+H^2} R,
\end{align}
where
\begin{align}
    E &= \sqrt{1+\tan^2(\tth)-\cos^2(\alpha_V) \left(\sin(\beta_V)-\cos(\beta_V)\tan(\tth)\right)^2} \\
    F &= \cos(\alpha_V) \left(\sin(\beta_V) - \cos(\beta_V)\tan(\tth) \right) \\
    G &= \frac{E -\tan(\tth) F}{1+\tan^2(\tth)} - \cos(\alpha_V) \cos(\beta_V) \\
    H &= \frac{F + \tan(\tth)E}{1+\tan^2(\tth)} - \cos(\alpha_V) \sin(\beta_V) \\
    x_o &= \frac{E -\tan(\tth) F}{1+\tan^2(\tth)} R \\
    y_o &= \frac{F + \tan(\tth)E}{1+\tan^2(\tth)} R.
\end{align}

Since $\alpha_V$, $\beta_V$, and $\tth$ do not change for different $R$, we can factor out $R$ from $\ell_V(\tth)$.
Denote $\ell_V(\tth) = D \cdot d\ell_V(\tth)$, using \eq{correction-term}, we have 
\begin{align}
    c_{v_e} &= \frac{N_V}{\sum_V e^{-\mu D \cdot d\ell_V(\tth)}}.
\end{align}

This concludes the proof that the same \mud would give the same \cve and we can consider it as an independent variable.

\section{Fast calculation of the absorption correction using a polynomial approach}
\label{sec:fast_calc}

Part of the popularity of \getx is the simplicity and speed of applying \adhoc corrections to the diffraction data that require little user input and still result in reliable PDFs \cite{billingeRobustAdHoc2013}. 
For longer wavelength x-rays where absorption is important we would like a similarly quick and straightforward absorption correction that can be incorporated into the \getx workflow. 

Here we describe a fast approximate calculator of \cve for a given \mud.
Since our recommended protocols for data acquisition are to measure samples with moderate absorption, i.e., $0.5 \leq \mud \leq 7$, we confine the estimates to this range. 
As a start, we explore estimating \cve through an interpolative reconstruction approach using polynomials, given a known value for \mud. 
Our strategy is to build a relatively sparse database of \cve curves computed by brute force, and to find curves for \mud values in between using interpolation of polynomials. 
The $1/\cve$ curves vary less strongly than \cve, with all values falling between 0 and 1, and we found the interpolation works better on  $1/\cve$ than on \cve itself.

\begin{figure}
    \centering
    \includegraphics[width= 0.65 \textwidth]{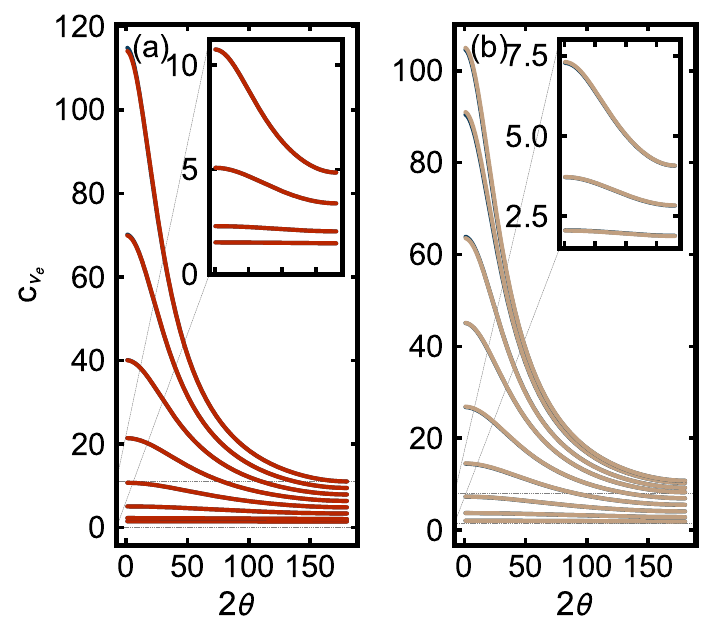}
    \caption{Comparison of brute-force and fast calculation of \cve.
    (a) Polynomial fits (red) and brute-force-calculated curves (blue, almost invisible) for \mud values between 0.5 and 7 from our database of curves. (b) Polynomial curves found by interpolation from the database curves (brown), and brute-force curves computed for the same \mud values (blue, almost invisible), for various \mud's randomly chosen between \mud's in the sparse dataset. In both panels, the insets show curves for smaller \mud's on an expanded scale. From bottom to top in (a) \mud = 0.5, 1, 2, 3, 4, 5, 6, and 7, and in (b) \mud = 0.8637, 1.5973, 2.47, 3.42, 4.344, 5.2, 5.824, 6.5, and 6.8.}
    \label{fig:cve_from_poly_fit}
\end{figure}

We started by creating a sparse dataset of $1/\cve$ curves for $\mud = 0.5$, 1, 2, 3, 4, 5, 6, and 7 over a range $1 \le \tth \le 180 \degree$ on a grid of spacing of $\Delta\tth = 0.1\degree$ using the brute-force approach. 
We then carried out a fit of a polynomial  to each of these curves.
After experimenting with using different polynomial degrees, we found that a sixth-degree polynomial provided reliable reconstructions.
The reconstructed curves are very close to the original ones as can be seen in \fig{cve_from_poly_fit}(a) which shows a comparison on the brute force and polynomial fits for the \cve curves chosen above.
The fits are very good, as the blue brute-force curves are barely visible. Small errors are seen clustered around low \tth, with  maximum percentage error of around 0.68\% for $\mud=7$ at $\tth = 1 \degree$. 
The maximum percentage errors for $\mud=4$, 5, and 6 also occurred at $\tth=1 \degree$, with values of 0.03\%, 0.12\%, 0.32\%, respectively.
For $\mud \leq 3$, the errors are almost negligible, with the maximal percentage error much less than $\leq 10^{-2}$\%.
The polynomial coefficients appropriate to each \mud are then stored in a database.

We then need a method to find the \cve curves for any \mud that may lie between those in our sparse database.
To interpolate between curves of different \mud, we used quadratic interpolation for the polynomial coefficients. 
We also experimented with linear and cubic interpolations: the former is not accurate enough and the latter tends to over-fit, settling on quadratic interpolation as the best approach.
The full \cve curve is then reconstructed using the resulting polynomial function.
The accuracy of this approach is illustrated in \fig{cve_from_poly_fit} (b). 
In the figure the blue curves are computed for intermediate \mud values using the full brute force calculation, and the light brown curve is the one estimated using the polynomial interpolation approach.
Since the maximum percentage errors are $<1.1$\% for all \mud's, we conclude that our model provides highly accurate estimations for \cve across various \mud values within our chosen range of $0.5 \leq \mud \leq 7$, while significantly reducing the computation time from hours to seconds.

\section{Estimation of \mud using model fitting of $z$-scans}
\label{sec:fit_mud_si}

Here we develop a model for estimating \mud experimentally from \zscan data.
We define the $x$ direction as the direction of travel of the x-ray beam and $z$ to be the vertical direction perpendicular to the capillary (see \fig{zscan_scheme}).  
We place the origin at the center of the x-ray beam.
Let $R=D/2$ be the radius of the capillary cross-section, and $h$ to be the height of the x-ray beam.
Finally, we define $z_0$ to be the height of the center of the capillary.  
$z_0$ could take any value without loss of generality but for convenience we try and set $z_0\sim 0$.  This offset will be refined in the fit.

The path length through the sample of an x-ray ray at the height, $z$ is then given by
\begin{equation}
l = \begin{cases} 2\sqrt{R^2-(z-z_0)^2} & \text{for } |z-z_0| \leq R \\
0 & \text{otherwise}.
\end{cases}
\end{equation}

We define $I_0$ as the intensity of the unattenuated x-ray beam (assumed to have a uniform cross-section).
We found that, in general, $I_0$ had a small, approximately linear, dependence on $z$, so we also define a coefficient, $m$, to account for this to obtain the best fits.
In addition, we assume $z$ is sampled at discrete intervals with a constant step size $\Delta z$.

We therefore define the intensity function $I[z]$ as
\begin{align}
I[z] = \begin{cases} 
(I_0 - mz) \cdot e^{-2\mu \sqrt{R^2-(z-z_0)^2}} & \text{for } |z-z_0| \leq R \\
I_0 - mz & \text{otherwise}.
\end{cases}
\label{eq:intensity_function}
\end{align}

The normalized kernel function $K[z]$ to account for the finite width beam is given by
\begin{align}
K[z] = \frac{1}{C} \begin{cases} 
1 & \text{for } |z| \leq h/2 \\
0 & \text{otherwise},
\end{cases}
\label{eq:kernel_function}
\end{align}
where $C$ is a constant such that $\sum_z K[z]=1$. Therefore, $C$ is the number of $x$ such that $|z| \leq h/2$, and $C = 2 \lfloor{\frac{h}{2\Delta t}}\rfloor + 1$.

To fit the measured curves we need to convolute $I[z]$ with $K[z]$, denoted as $(I \circledast K)[z]$, given by
\begin{align}
    (I \circledast K)[z] &= \sum_t I[t] \cdot K[z-t] \\
    &= \sum_t I[t] \begin{cases}
    \frac{1}{C} & \text{for } |z-t| \leq h/2 \\
    0 & \text{otherwise}
    \end{cases} \\
    &= \frac{1}{C} \sum_{t:|z-t|\leq h/2} I[t] \\
    &= \frac{1}{C} \sum_{t:|z-t|\leq h/2} \begin{cases}
    (I_0 - mt) \cdot e^{-2\mu \sqrt{R^2-(t-z_0)^2}} & \text{for } |t-z_0| \leq R \\
    I_0 - mt & \text{otherwise}
    \end{cases} \\
    &= \frac{1}{C} \sum_{k=1}^{C} \begin{cases}
    (I_0 - mt_k) \cdot e^{-2\mu \sqrt{R^2-(t_k-z_0)^2}} & \text{for } |t_k-z_0| \leq R \\
    I_0 - mt_k & \text{otherwise},
    \end{cases}
    \label{eq:convolution_function}
\end{align}
where $t_k \in [z-h/2, z+h/2]$ with $t_{i+1}=t_i+\Delta t$ for $i=1,2,...,C-1$.

As seen from \eq{convolution_function}, the convolution at each $z$ is essentially an average of $I[t_k]$ for each $t_k \in [z-h/2, z+h/2]$. As $h$ increases, the convolution deviates more significantly from the intensity function in \eq{intensity_function} and as $h \rightarrow 0$, $C \rightarrow 1$ meaning that the number of $t_k$ we are averaging approaches $1$, in which case, \eq{convolution_function} is just the unconvoluted intensity, \eq{intensity_function}.

We have developed a user-friendly fitting program in \utils \cite{farrowDiffpyutils2024} that accurately and efficiently extracts \mud by optimizing the six parameters: $\mu$, $D$, $h$, $I_0$, $z_0$, and $m$. 
The function only takes in a set of experimental $z$ and $I$ values and minimizes the sum of squared residuals between the experimental data and the model convolution. The convolution is computed using SciPy signal's convolve function with same mode, based on the intensity and kernel functions defined in \eq{intensity_function} and \eq{kernel_function}. 
Boundary effects occur because the kernel only partially overlaps with the intensity signal at the edges, resulting in incomplete averaging. To address this, we extend the $z$ values for both intensity and kernel functions to ensure full coverage for the given $z$ values. The same mode also introduces a potential shift in convolution as it crops the result to match the size of $z$. As a result, we recenter the kernel around $z-\bar{z}$, where $\bar{z}$ is the average of $z$, aligning the intensity with the convolution.
To optimize the search for the global minimum, we use SciPy's dual annealing method, a stochastic global optimization algorithm. This approach provides a robust solution to find the global minimum. By combining the theoretical results from \eq{intensity_function}, \eq{kernel_function}, and \eq{convolution_function} with the fitting program, we achieve a more accurate estimation of \mud.

\section{Estimation of \mud using a theoretical database}
\label{sec:theoretical_mud_si}
The process for estimating the theoretical linear attenuation coefficient $\mu$ involves using the XrayDB database \citexraydb, which can then be used to compute \mud by multiplying with the capillary diameter $D$.
The inputs required for this calculation are the sample composition, the x-ray energy, and either a measured mass density, $\rho_s$, for the sample, or the packing fraction $f$, where $\rho_s = \rho_m \cdot f$, where $\rho_m$ is the mass density of fully dense sample that can be estimated from the atomic structure if it is known.

To calculate $\mu$, we first compute the mass attenuation coefficient $\left(\frac{\mu}{\rho}\right)_e$ for each element $e$ in the sample, in cm$^2$/g. A mass-weighted sum approach is then applied to determine the total mass attenuation coefficient for the mixture. 
Let $E$ be the set of sample elements in the sample, $f_e$ be the number of atoms per element $e$, and $A_e$ be its atomic mass in g/mol. Then the mass contribution of each element $e$ is given by $m_e = f_e \cdot A_e$, from which the total mass attenuation coefficient $\left(\frac{\mu}{\rho}\right)_{\text{tol}}$ and total mass $m_{\text{tol}}$ are computed as
\begin{align}
    \left(\frac{\mu}{\rho}\right)_{\text{tol}} &= \sum_{e \in E} \left(\frac{\mu}{\rho}\right)_e \cdot m_e \\
    m_{\text{tol}} &= \sum_{e \in E} m_e.
\end{align}
The final $\mu$ value is then obtained by normalizing $\left(\frac{\mu}{\rho}\right)_{\text{tol}}$ by $m_{\text{tol}}$ and scaling by the sample mass density $\rho_s$:
\begin{align}
    \mu = \rho_s \cdot \frac{\left(\frac{\mu}{\rho}\right)_{\text{tol}}}{m_{\text{tol}}}.
\end{align}

\section{Selecting \mude in different situations}
\label{sec:select_mude}

In the main paper we assessed the effect of different \cve corrections on refined structural parameters from models.
Here, for completeness, we present more complete results of this study by taking the optimal \mud corrections but applying them to data from different thickness \ce{CeO2} samples with different degrees of absorption.

In Table \ref{table:CeO2_best_corrections} we show the refined parameters for synchrotron, uncorrected and best-corrected data for all \ce{CeO2} datasets, with additional \mud's tested listed in Tables \ref{table:CeO2_635_fits_SI}, \ref{table:CeO2_813_fits_SI}, and \ref{table:CeO2_1024_fits_SI}. 
We find that \mude depends on different experimental conditions, specifically the relative x-ray beam height $h$ and capillary diameter~$D$. 
For data collected with $D \approx 0.8 * h$ (IDs 1.024~mm and 0.813~mm here, with $h=1.2$~mm and 1.0~mm, respectively), the \mude's, 6.5 and 4, are close to their theoretical values, 5.824 and 4.554, respectively.  
For data with $D \ll h$ (ID $=0.635$~mm with $h=1$~mm), $\mude = 1.938$ which is much smaller than $\mudmfour =4.229$. 
Notably, this value for \mude was obtained from the $z$-scan absorption measurement that was made with the slit and detector in the experimental condition.

If it is possible to measure sample absorption using a \zscan and using the fitting program in our \labpdfproc software to extract \mud, it is recommended to do this with the slits for the \zscan in the same settings as was used to collect the experimental data.
However, if it is not possible, \mude can be approximated by scaling the theoretical value \mudmfour by $D/h$ to get an approximately correct value for the effective \mud to use in the \cve calculation.
Future investigations with more data will be necessary to fully prove this equation.

\begin{table}
\begin{center}
\caption{Results of fittings for \ce{CeO2} synchrotron, uncorrected, and best-corrected data, over $r_{\min} = 1.0$ and $r_{\max} = 40.0$. The uncorrected data are listed with their IDs (1.024~mm, 0.813~mm, and 0.635~mm), and the best-corrected data are labeled with the ID followed by ``corrected".}
\label{table:CeO2_best_corrections}
\begin{adjustbox}{scale=0.8}
\begin{tabular}{llllllll}
\toprule
Parameter & $1.024$~mm & $0.813$~mm & $0.635$~mm & $1.024$~mm  & $0.813$~mm  & $0.635$~mm  & synchrotron \\
~ & ~ & ~ & ~ & corrected & corrected & corrected & ~ \\
\midrule
       $s_1$ &                                          0.2191(9) &                    0.3060(12) &                     0.2923(9) &                                       0.3711(11) &                                    0.4199(13) &                                         0.4331(11) & 0.37434(21) \\
    \qdamp &                                          0.03186(25) &                   0.02971(23) &                   0.02839(19) &                                      0.02448(24) &                                   0.02650(21) &                                        0.02655(17) & 0.02386(4) \\
   \qbroad &                                            0.0220(5) &                     0.0241(4) &                   0.02405(34) &                                        0.0373(4) &                                     0.0308(4) &                                        0.02841(31) & 0.01814(7) \\
   $\delta_2$ &                                               9.92(28) &                      9.70(17) &                      9.68(12) &                                         7.14(11) &                                       7.18(9) &                                            7.18(7) & 9.07(4) \\
        $a$ &                                          5.40240(5) &                    5.40233(5) &                    5.40398(4) &                                       5.39937(6) &                                    5.40071(5) &                                         5.40291(4) & 5.414232(7) \\
   Ce(\uiso) &                                        0.002403(24) &                  0.002751(24) &                  0.003098(21) &                                     0.003388(31) &                                  0.003474(28) &                                       0.003629(23) & 0.003571(5)  \\
   O(\uiso) &                                           0.0990(24) &                    0.0765(14) &                     0.0670(9) &                                        0.0405(5) &                                     0.0500(6) &                                          0.0542(6) & 0.04120(10) \\
     $Q_{\max}$ &                                                         16.6 &                          16.6 &                          16.6 &                                             16.6 &                                          16.6 &                                               16.6 & 30.0 \\
     grid &                                                 0.189253 &                      0.189253 &                      0.189253 &                                         0.189253 &                                      0.189253 &                                           0.189253 & 0.10472 \\
       \rw &                                                 0.718991 &                      0.581325 &                      0.507961 &                                         0.390974 &                                      0.354827 &                                            0.34172 & 0.160823 \\
  $\chi_{red}^2$ &                                            1173.593401 &                    758.117681 &                    864.835553 &                                       390.094422 &                                    330.294145 &                                         434.106299 & 961.528958\\
\bottomrule
\end{tabular}
\end{adjustbox}
\end{center}
\end{table}

\section{Refinement across extended \mud values}
\label{sec:low_and_high_mud}

For a low \mud $<2$, the uncorrected lab data already provides a reasonably good fit (low \rw). This is demonstrated in Table \ref{table:ZrO2_fits} for \ce{ZrO2} packed in a 1.024~mm capillary, where the theoretical \mud is $\sim 1.5$. 
We have applied two corrections, corresponding to the largest and smallest experimental \mud's computed from Method 2, using $h=0.05$~mm and 1.2~mm, respectively, with a reduced channel. 
To avoid overfitting, isotropic ADPs were used for zirconium and oxygen.

Overall, we find that the uncorrected lab data provides results comparable to the synchrotron data. 
As with the ceria data described in the main paper, the lattice parameters of the synchrotron data are slightly overestimated.
In ceria, the ADP on the metal site was most important.
Here the difference in scattering power between oxygen and Zr is smaller and the oxygen ADPs are more significant.
We find that the ADPs of both the metal and oxygen increase with increasing absorption correction as expected following the arguments made in the main paper.
However, even for the uncorrected \ce{ZrO2}, the ADPs of the lab data are higher than those from the synchrotron data.
We are not sure why this is the case.  The refinements are quite reproducible between the synchrotron and lab data, considering the significant differences between the resolutions and \q-ranges of the different measurements, but do not allow a clear differentiation between the different absorption corrections.

\begin{figure}[H]
   \centering
   \includegraphics[width=\linewidth]{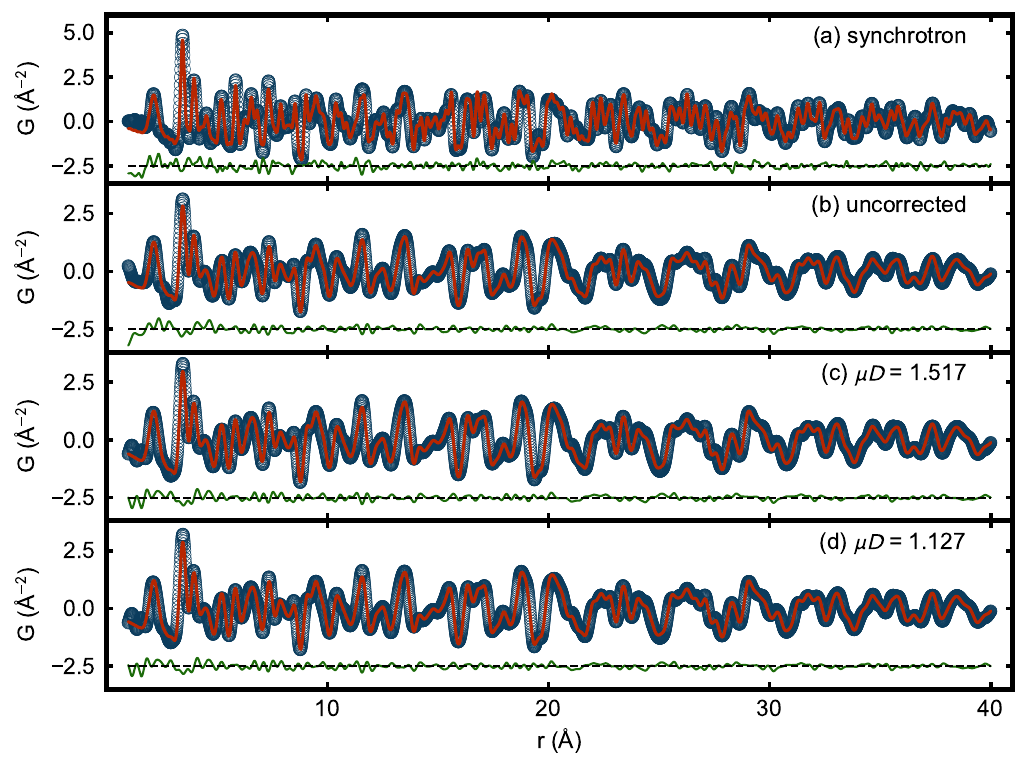}
   \caption{Results of \ce{ZrO2} synchrotron, uncorrected, and corrected laboratory data over $r_{\min} = 1.0$ and $r_{\max} = 40.0$. The \mud's reported inside the panels are the values used for correction. The measured PDFs are the blue circles, the best-fit PDFs are the red lines, and the residuals shown are green. 
   }
   \label{fig:ZrO2_PDFs}
\end{figure}

The PDFs of the synchrotron and laboratory data are plotted in \fig{ZrO2_PDFs}. 
The most significant difference is that the lab data have slightly broader peaks compared to the synchrotron data due to the lower \qmax, but all the fits are of high quality.

\begin{table}
\begin{center}
\caption{Results of fittings for \ce{ZrO2} data (ID=1.024~mm) for synchrotron, uncorrected, and corrected lab data, over $r_{\min} = 1.0$ and $r_{\max} = 40.0$. The \mud's are the values applied for correction.
}
\label{table:ZrO2_fits}
\begin{tabular}{lllll}
\toprule
Parameter & synchrotron & uncorrected & \mud = 1.127 & \mud = 1.517 \\
\midrule
       $s_1$ &                       0.49519(28) &                            0.5314(27) &                                                                                  0.5771(28) & 0.6077(29) \\
    \qdamp &                        0.02282(5) &                             0.0201(5) &                                                                                    0.0218(4) & 0.0215(4) \\
   \qbroad &                       0.03182(13) &                             0.0329(9) &                                          0.0298(8) &                                          0.0294(8) \\
   $\delta_2$ &                         1.294(17) &                              1.94(10) &                                            1.85(9) &                                            1.80(9) \\
        $a$ &                      5.148597(23) &                           5.14610(30) &                                                                    5.14606(29) & 5.14622(28) \\
        $b$ &                      5.211724(23) &                           5.20949(34) &                                                                              5.20991(32) & 5.21009(31) \\
        $c$ &                      5.317103(24) &                           5.31141(35) &                                                                               5.31192(34) & 5.31218(33) \\
     $\beta$ &                        99.2208(4) &                             99.205(6) &                                          99.201(6) &                                          99.201(5) \\
  Zr(\uiso) &                     0.0010703(25) &                            0.00451(8) &                                                                               0.00507(9) & 0.00532(9) \\
   O(\uiso) &                       0.03721(12) &                            0.0397(10) &                                         0.0438(10) &                                         0.0445(10) \\
     \qmax &                              30.0 &                                  16.6 &                                               16.6 &                                               16.6 \\
     grid &                           0.10472 &                              0.189253 &                                           0.189253 &                                           0.189253 \\
       \rw &                          0.218742 &                              0.224071 &                                                                                     0.197061 & 0.191054 \\
  $\chi_{red}^2$ &                       1298.813481 &                             37.001624 &                                                                                    32.546912 & 33.036836 \\
\bottomrule
\end{tabular}

\end{center}
\end{table}

\begin{figure}
    \centering
    \includegraphics[width= 0.7\textwidth]{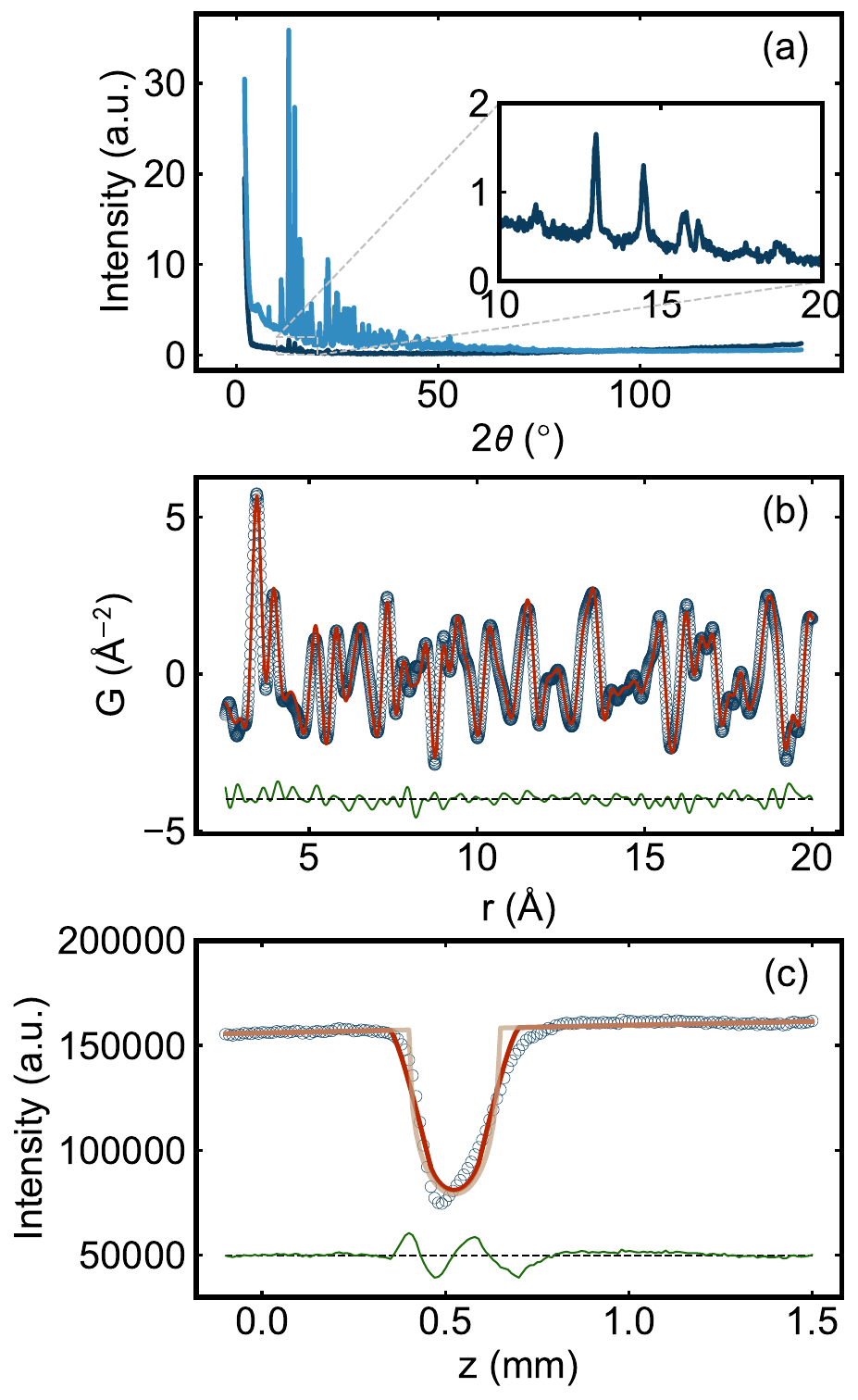}
    \caption{
    XRD, PDF, and \zscan analysis of \ce{HfO2} data.
    (a) XRD comparison of \ce{HfO2} measured with a capillary (dark blue) and wire (light blue), with an inset showing an expanded portion of the capillary data highlighting its only signals.
    (b) PDF fit for wire data, showing the measured PDF (blue), the best-fit PDF (red), and residuals (green), fitted between $r_{\min} = 1.0$ and $r_{\max} = 20.0$.
    (c) \zscan fit for wire data, with the original intensity (blue), the fit (red), the unconvoluted intensity (brown), and residuals (green). 
    }
    \label{fig:HfO2_plot}
\end{figure}

We now turn to the case of very high \mud by considering data from the \ce{HfO2} sample. 
As discussed in the main paper, the significant absorption of the sample results in a strong suppression of the diffraction signal, resulting in a very low signal/noise ratio. 
This is evident by looking at the dark blue curve in \fig{HfO2_plot}(a), which shows the XRD data for a measurement of \ce{HfO2} using a capillary of 0.635~mm, where the theoretical \mud is just above eight. As shown in the inset, only a few weak signals are observed, and the curve remains relatively flat until the end, where there is a strong upturn. 
While the absorption correction would adjust signal amplitudes, it amplifies both the signals and the noise by the same amount, leaving the signal/noise ratio unaffected. 
As a result, the data is functionally useless for PDF analysis. 

\begin{figure}
    \centering
    \includegraphics[width=0.8\linewidth]{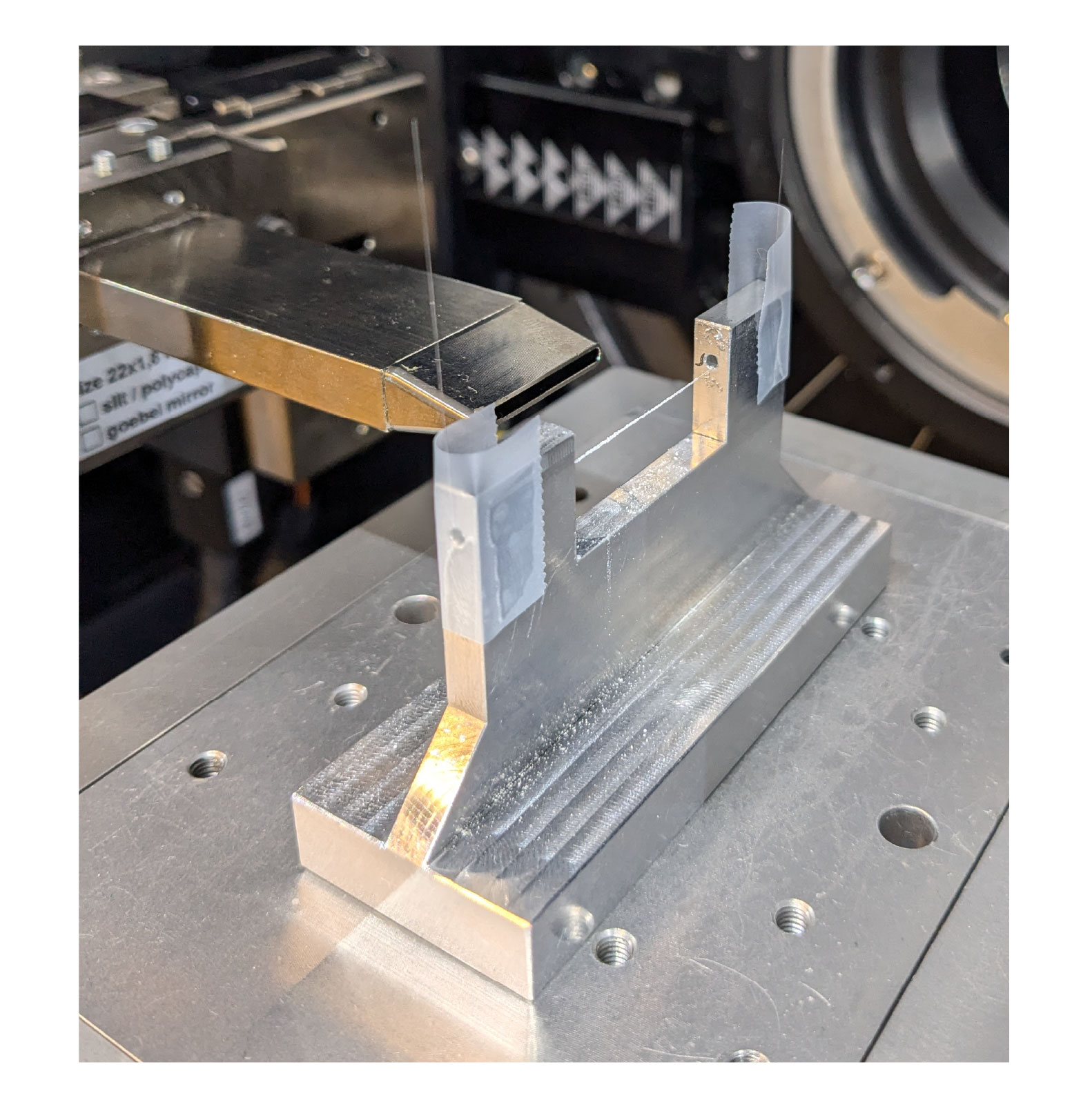}
    \caption{Photograph of the wire coated with the powder of HfO$_2$. The wire was coated with some grease, then, the hafnia powder was dispersed over the coated area. }
    \label{fig:hfo2_wire_image}
\end{figure}

It is possible to load the sample into a thinner capillary if one is available.
However,  a simple workaround is to take a glass or plastic wire and rub some grease and powder on the outside of the wire (see Figure \ref{fig:hfo2_wire_image}). This approach is very effective if the powder scatters well enough, which is typically the case for elements with large x-ray cross-sections. 
In \fig{HfO2_plot} (a), the XRD data for the wire is shown in light blue. 
Comparing it to the dark blue curve (representing \ce{HfO2} in the 0.635~mm capillary), the wire data shows a significant improvement in the signal/noise ratio, with a reduced upturn at the end, indicating much better data quality. 
To evaluate the effectiveness further, \fig{HfO2_plot} (b) shows the PDF obtained from the uncorrected wire diffraction pattern, fitted between $r_{\min} = 1.0$ and $r_{\max} = 20.0$~\AA. The refined parameters are provided in Table \ref{table:HfO2_fitting_table}. The small residuals of the fit confirm the high data quality, making the wire data the preferred choice.

Finally, we present a method to estimate the theoretical \mud for wire data, which is useful for determining the effective \mud to use to apply for the absorption correction.
Direct theoretical estimation is impossible since we are using a wire instead of a capillary, meaning that we do not know the powder density and exact diameter of the wire and powder together. 
As a result, we can use a \zscan to obtain experimental values, using, for instance, a small x-ray beam height of 0.1~mm with a reduced channel. 
This returns an experimental \mud value close to the theoretical one. 
\fig{HfO2_plot} (c) shows the \zscan fit using Method 2 with the experimental settings mentioned above. Although the fitted red curve shows some deviation from the original blue intensity data, making it less accurate, we find it still reliable, as verified through both Method 1 (\mudmone $= 0.777$) and Method 2 (\mudmtwo $= 0.691$).

\begin{table}
\begin{center}
\caption{Results of the refinement for uncorrected \ce{HfO2} wire data over $r_{\min} = 1.0$ and $r_{\max} = 20.0$.}
\label{table:HfO2_fitting_table}
\begin{tabular}{lcccccc}
\toprule
Parameter & \multicolumn{1}{c}{Uncorrected} \\
\midrule
       $s_1$ & 0.667(13) \\
    \qdamp & 0.0480(22) \\
   \qbroad & 0.0(5) \\
   $\delta_2$ & 6.47(24) \\
        $a$ & 5.1106(11) \\
        $b$ & 5.1780(15) \\
        $c$ & 5.2932(10) \\
     $\beta$ & 99.544(23) \\
    U11\_0 & 0.0094(4) \\
    U22\_0 & 0.0091(5) \\
    U33\_0 & 0.00808(32) \\
    U12\_0 & 0.0092(4) \\
    U13\_0 & 0.0088(4) \\
    U23\_0 & 0.0087(4) \\
    U11\_4 & 0.0083(25) \\
    U22\_4 & 0.0101(26) \\
    U33\_4 & 0.0113(25) \\
    U12\_4 & 0.0094(25) \\
    U13\_4 & 0.0099(25) \\
    U23\_4 & 0.0106(25) \\
    U11\_8 & 0.0109(28) \\
    U22\_8 & 0.0096(25) \\
    U33\_8 & 0.0100(27) \\
    U12\_8 & 0.0104(26) \\
    U13\_8 & 0.0103(27) \\
    U23\_8 & 0.0104(26) \\
      x\_0 & 0.2747(4) \\
      y\_0 & 0.46602(31) \\
      z\_0 & 0.7046(4) \\
      x\_4 & 0.1041(19) \\
      y\_4 & 0.2126(19) \\
      z\_4 & 0.8875(19) \\
      x\_8 & 0.4889(21) \\
      y\_8 & 0.7553(18) \\
      z\_8 & 0.9863(22) \\
     $Q_{\max}$ & 16.6 \\
     grid & 0.189253 \\
       \rw & 0.366525 \\
  $\chi_{\text{red}}^2$ & 41.722268 \\
\bottomrule
\end{tabular}
\end{center}
\end{table}

\section{Additional Tables and Figures}
\label{sec:full_mud_table}

\begin{table}[H]
    \caption{List of \mud's computed using the four different methods described in the texts, labeled as Method 1, Method 2, Method 3, and Method 4.
    }
    \begin{center}
    \begin{tabular}{lcccccccc}
    \toprule
    Sample & ID & Density & $h$ & Channel & Method 1 & Method 2 & Method 3 & Method 4 \\
    & (mm) & ($g/cm^3$) & (mm) & & & & \\
    \midrule
    \ce{ZrO2} & 0.635 & 1.009 & 0.10 & Closed & 0.615 & 0.538 & 0.517 & 0.795 \\
    ~ & 0.813 & 0.856 & 0.10 & Closed & 0.711 & 0.592 & 0.592 & 0.864 \\
    ~ & 1.024 & 1.122 & 0.05 & Closed & 1.555 & 1.517 & 1.529 & 1.426 \\
    ~ & ~ & ~ & 0.20 & Closed & 1.399 & 1.398 & 1.421 & ~ \\
    ~ & ~ & ~ & 0.60 & Closed & 1.120 & 1.120 & 1.206 & ~ \\
    ~ & ~ & ~ & 1.20 & Closed & 1.142 & 1.127 & 1.200 & ~ \\
    \ce{CeO2} & 0.635 & 1.706 & 0.05 & Closed & 3.741 & 3.661 & 3.624 & 4.229 \\
    ~ & ~ & ~ & 0.10 & Closed & 3.714 & 3.666 & 3.671 & ~ \\
    ~ & ~ & ~ & 0.20 & Closed & 3.519 & 3.489 & 3.519 & ~ \\
    ~ & ~ & ~ & 0.60 & Closed & 3.108 & 3.108 & 3.430 & ~ \\
    ~ & ~ & ~ & 1.00 & Closed & 3.055 & 3.054 & 3.384 & ~ \\
    ~ & ~ & ~ & 0.05 & Open & 3.280 & 3.245 & 3.208 & ~ \\
    ~ & ~ & ~ & 0.20 & Open & 3.001 & 3.011 & 3.001 & ~ \\
    ~ & ~ & ~ & 0.60 & Open & 1.971 & 2.005 & 2.098 & ~ \\
    ~ & ~ & ~ & 1.00 & Open & 1.912 & 1.938 & 2.016 & ~ \\
    ~ & 0.813 & 1.435 & 0.10 & Closed & 4.543 & 4.465 & 4.509 & 4.554 \\
    ~ & 1.024 & 1.457 & 0.10 & Closed & 5.419 & 5.328 & 5.391 & 5.824 \\
    \ce{HfO2} & 0.635 & 1.741 & -- & -- & -- & -- & -- & 8.168 \\
    ~ & 0.813 & 1.963 & 0.10 & Open & 5.894 & 5.575 & 5.570 & 11.79 \\
    ~ & Wire
    & -- & 0.10 & Closed & 0.777 & 0.691 & -- & -- \\
    \bottomrule
    \end{tabular}
    \label{tab:mud_list}
    \end{center}
\end{table}

\begin{table}
    \caption{List of fitted and theoretical or given physical parameters $\mu$, $D$, and $h$. $\mu_{\text{theoretical}}$ is the theoretical $\mu$ computed from XrayDB, $D_{\text{given}}$ is the given inner diameter (ID) of each capillary, $h_{\text{given}}$ is the given x-ray beam height. $\mu_{\text{fit}}$, $D_{\text{fit}}$, and $h_{\text{fit}}$ are the fitted parameters computed from Method 2 for each $z$-scan data.}
    \begin{center}
    \begin{tabular}{lcccccccc}
    \toprule
    Sample & $D_{\text{given}}$ & Density & $h_{\text{given}}$ & Channel & $\mu_{\text{theoretical}}$ & $\mu_{\text{fit}}$ & $D_{\text{fit}}$ & $h_{\text{fit}}$ \\
    & (mm) & ($g/cm^3$) & (mm) & & (mm$^{-1}$) & (mm$^{-1}$) & (mm) & (mm) \\
    \midrule
    \ce{ZrO2} & 0.635 & 1.009 & 0.10 & Closed & 1.252 & 0.814 & 0.660 & 0.095 \\
    ~ & 0.813 & 0.856 & 0.10 & Closed & 1.062 & 0.728 & 0.813 & 0.042 \\
    ~ & 1.024 & 1.122 & 0.05 & Closed & 1.392 & 1.493 & 1.016 & 0.067 \\
    ~ & ~ & ~ & 0.20 & Closed & ~ & 1.387 & 1.008 & 0.167 \\
    ~ & ~ & ~ & 0.60 & Closed & ~ & 1.184 & 0.948 & 0.290 \\
    ~ & ~ & ~ & 1.20 & Closed & ~ & 1.172 & 0.962 & 0.295 \\
    \ce{CeO2} & 0.635 & 1.706 & 0.05 & Closed & 6.659 & 5.707 & 0.642 & 0.073 \\
    ~ & ~ & ~ & 0.10 & Closed & ~ & 5.781 & 0.634 & 0.075 \\
    ~ & ~ & ~ & 0.20 & Closed & ~ & 5.542 & 0.630 & 0.148 \\
    ~ & ~ & ~ & 0.60 & Closed & ~ & 5.401 & 0.575 & 0.276 \\
    ~ & ~ & ~ & 1.00 & Closed & ~ & 5.330 & 0.583 & 0.280 \\
    ~ & ~ & ~ & 0.05 & Open & ~ & 5.052 & 0.642 & 0.092 \\
    ~ & ~ & ~ & 0.20 & Open & ~ & 4.725 & 0.637 & 0.176 \\
    ~ & ~ & ~ & 0.60 & Open & ~ & 3.304 & 0.607 & 0.322 \\
    ~ & ~ & ~ & 1.00 & Open & ~ & 3.174 & 0.610 & 0.332 \\
    ~ & 0.813 & 1.435 & 0.10 & Closed & 5.601 & 5.546 & 0.805 & 0.672 \\
    ~ & 1.024 & 1.457 & 0.10 & Closed & 5.687 & 5.265 & 1.012 & 0.073 \\
    \ce{HfO2} & 0.635 & 1.741 & -- & -- & 12.86 & -- & -- & -- \\
    ~ & 0.813 & 1.963 & 0.10 & Open & 14.50 & 6.851 & 0.814 & 0.083 \\
    ~ & Wire
    & -- & 0.10 & Closed & -- & 2.804 & 0.247 & 0.104 \\
    \bottomrule
    \end{tabular}
    \label{tab:fitted_parameters}
    \end{center}
\end{table}

\begin{table}
\begin{center}
\caption{Results of the refinement for \ce{CeO2} data (ID=0.635~mm) for synchrotron, uncorrected, and corrected data with different \mud's, over $r_{\min} = 1.0$ and $r_{\max} = 40.0$. The \mud's are the values used for each correction.}
\label{table:CeO2_635_fits_SI}
\begin{adjustbox}{angle=-90, scale=0.8}
\begin{tabular}{llllllllll}
\toprule
Parameter & synchrotron & uncorrected & \mud = 1.912 & \mud = 1.971 & \mud = 2.005 & \mud = 3.054 & \mud = 3.489 & \mud = 3.666 & \mud = 3.741 \\
\midrule
       $s_1$ &                       0.37434(21) &                     0.2923(9) &                                         0.4288(11) &                                         0.4305(11) &                                         0.4314(11) &                                         0.4534(10) &                                         0.4547(10) &                                         0.4540(10) &                                         0.4539(10) \\
    \qdamp &                        0.02386(4) &                   0.02839(19) &                                        0.02627(18) &                                        0.02621(18) &                                        0.02618(17) &                                        0.02525(17) &                                        0.02463(16) &                                        0.02444(16) &                                        0.02435(16) \\
   \qbroad &                        0.01814(7) &                   0.02404(34) &                                        0.02720(32) &                                        0.02725(32) &                                        0.02728(32) &                                        0.02813(29) &                                        0.02782(27) &                                        0.02769(27) &                                        0.02759(27) \\
   $\delta_2$ &                           9.07(4) &                      9.67(12) &                                            8.26(9) &                                            8.22(9) &                                            8.20(9) &                                            7.15(7) &                                            7.14(6) &                                            7.14(6) &                                            7.13(6) \\
        $a$ &                       5.414232(7) &                    5.40398(4) &                                         5.40301(4) &                                         5.40298(4) &                                         5.40297(4) &                                         5.40255(4) &                                         5.40242(4) &                                         5.40239(4) &                                         5.40237(4) \\
   Ce(\uiso) &                       0.003571(5) &                  0.003099(21) &                                       0.003712(24) &                                       0.003735(24) &                                       0.003749(24) &                                       0.004317(26) &                                       0.004734(27) &                                       0.004878(27) &                                       0.004959(27) \\
   O(\uiso) &                       0.04120(10) &                     0.0670(9) &                                          0.0560(6) &                                          0.0558(6) &                                          0.0556(6) &                                          0.0519(5) &                                          0.0516(5) &                                          0.0517(4) &                                          0.0517(4) \\
     \qmax &                              30.0 &                          16.6 &                                               16.6 &                                               16.6 &                                               16.6 &                                               16.6 &                                               16.6 &                                               16.6 &                                               16.6 \\
     grid &                           0.10472 &                      0.189253 &                                           0.189253 &                                           0.189253 &                                           0.189253 &                                           0.189253 &                                           0.189253 &                                           0.189253 &                                           0.189253 \\
       \rw &                          0.160823 &                      0.507961 &                                           0.344361 &                                           0.340531 &                                           0.338342 &                                           0.270047 &                                           0.244028 &                                           0.237034 &                                           0.233655 \\
  $\chi_{red}^2$ &                        961.528958 &                    864.835556 &                                         439.844024 &                                          432.38352 &                                         428.156929 &                                           309.2045 &                                         271.704349 &                                          262.39813 &                                         258.344267 \\
\bottomrule
\end{tabular}
\end{adjustbox}
\end{center}
\end{table}

\begin{table}
\begin{center}
\caption{Results of the refinement for \ce{CeO2} data (ID=0.813~mm) for synchrotron, uncorrected, and corrected data with different \mud's, over $r_{\min} = 1.0$ and $r_{\max} = 40.0$. The \mud's are the values used for each correction.}
\label{table:CeO2_813_fits_SI}
\begin{adjustbox}{angle=-90, scale=0.8}
\begin{tabular}{llllllllll}
\toprule
Parameter & synchrotron & uncorrected & \mud = 2.15 & \mud = 2.5 & \mud = 3 & \mud = 3.5 & \mud = 4 & \mud = 4.554 & \mud = 5 \\
\midrule
       $s_1$ &                       0.37434(21) &                    0.3060(12) &                                       0.3562(13) &                                      0.3708(13) &                                    0.3902(13) &                                      0.4065(13) &                                    0.4173(13) &                                         0.4269(12) &                                    0.4234(12) \\
    \qdamp &                        0.02386(4) &                   0.02971(23) &                                      0.02875(22) &                                     0.02842(22) &                                   0.02785(22) &                                     0.02713(21) &                                   0.02628(21) &                                        0.02513(21) &                                   0.02438(21) \\
   \qbroad &                        0.01814(7) &                     0.0241(4) &                                        0.0261(4) &                                       0.0266(4) &                                     0.0276(4) &                                       0.0287(4) &                                     0.0300(4) &                                          0.0316(4) &                                     0.0322(4) \\
   $\delta_2$ &                           9.07(4) &                      9.69(17) &                                         9.04(14) &                                        8.86(13) &                                      8.56(13) &                                        8.22(12) &                                      7.83(11) &                                           7.36(11) &                                       7.15(9) \\
        $a$ &                       5.414232(7) &                    5.40233(5) &                                       5.40187(5) &                                      5.40168(5) &                                    5.40141(5) &                                      5.40110(5) &                                    5.40079(5) &                                         5.40023(5) &                                    5.40019(5) \\
   Ce(\uiso) &                       0.003571(5) &                  0.002751(24) &                                     0.002970(26) &                                    0.003052(26) &                                  0.003188(27) &                                    0.003351(28) &                                  0.003533(29) &                                       0.003760(31) &                                  0.004033(31) \\
   O(\uiso) &                       0.04120(10) &                    0.0765(14) &                                       0.0650(10) &                                      0.0622(10) &                                     0.0582(8) &                                       0.0544(7) &                                     0.0511(7) &                                          0.0479(6) &                                     0.0464(5) \\
     \qmax &                              30.0 &                          16.6 &                                             16.6 &                                            16.6 &                                          16.6 &                                            16.6 &                                          16.6 &                                               16.6 &                                          16.6 \\
     grid &                           0.10472 &                      0.189253 &                                         0.189253 &                                        0.189253 &                                      0.189253 &                                        0.189253 &                                      0.189253 &                                           0.189253 &                                      0.189253 \\
       \rw &                          0.160823 &                      0.581325 &                                         0.493463 &                                        0.467803 &                                      0.429734 &                                        0.391046 &                                      0.355844 &                                           0.322514 &                                      0.298757 \\
  $\chi_{red}^2$ &                        961.528958 &                    758.117673 &                                       554.991541 &                                      506.666438 &                                    441.041522 &                                      381.046435 &                                    332.191765 &                                         293.028937 &                                    265.892271 \\
\bottomrule
\end{tabular}
\end{adjustbox}
\end{center}
\end{table}

\begin{table}
\begin{center}
\caption{Results of the refinement for \ce{CeO2} data (ID=1.024~mm) for synchrotron, uncorrected, and corrected data with different \mud's, over $r_{\min} = 1.0$ and $r_{\max} = 40.0$. The \mud's are the values used for each correction.}
\label{table:CeO2_1024_fits_SI}
\begin{adjustbox}{angle=-90, scale=0.8}
\begin{tabular}{llllllll}
\toprule
Parameter & synchrotron & uncorrected & \mud = 3 & \mud = 5.328 & \mud = 5.824 & \mud = 6 & \mud = 6.5 \\
\midrule
       $s_1$ &                       0.37434(21) &                      0.2191(9) &                                     0.2972(12) &                                         0.3693(12) &                                         0.3785(12) &                                     0.3748(12) &                                       0.3712(11) \\
    \qdamp &                        0.02386(4) &                    0.03186(25) &                                    0.03074(24) &                                        0.02737(23) &                                        0.02612(23) &                                    0.02593(23) &                                      0.02449(24) \\
   \qbroad &                        0.01814(7) &                      0.0220(5) &                                      0.0251(5) &                                          0.0320(4) &                                          0.0347(4) &                                      0.0353(4) &                                        0.0372(4) \\
   $\delta_2$ &                           9.07(4) &                       9.92(28) &                                       9.21(17) &                                           7.87(13) &                                           7.42(13) &                                       7.15(11) &                                         7.14(11) \\
        $a$ &                       5.414232(7) &                     5.40240(5) &                                     5.40160(5) &                                         5.40018(5) &                                         5.39959(6) &                                     5.39967(6) &                                       5.39937(6) \\
   Ce(\uiso) &                       0.003571(5) &                   0.002403(24) &                                   0.002671(25) &                                       0.003154(29) &                                       0.003216(30) &                                   0.003265(29) &                                     0.003389(31) \\
   O(\uiso) &                       0.04120(10) &                     0.0990(24) &                                     0.0696(13) &                                          0.0472(7) &                                          0.0435(6) &                                      0.0425(6) &                                        0.0405(5) \\
     \qmax &                              30.0 &                           16.6 &                                           16.6 &                                               16.6 &                                               16.6 &                                           16.6 &                                             16.6 \\
     grid &                           0.10472 &                       0.189253 &                                       0.189253 &                                           0.189253 &                                           0.189253 &                                       0.189253 &                                         0.189253 \\
       \rw &                          0.160823 &                       0.718991 &                                       0.596404 &                                           0.439131 &                                           0.417204 &                                       0.407956 &                                         0.390974 \\
  $\chi_{red}^2$ &                        961.528958 &                    1173.593401 &                                     753.497185 &                                         447.645936 &                                         419.475451 &                                     407.497971 &                                       390.094511 \\
\bottomrule
\end{tabular}
\end{adjustbox}
\end{center}
\end{table}

\begin{figure}
    \centering
    \includegraphics[width=0.5\linewidth]{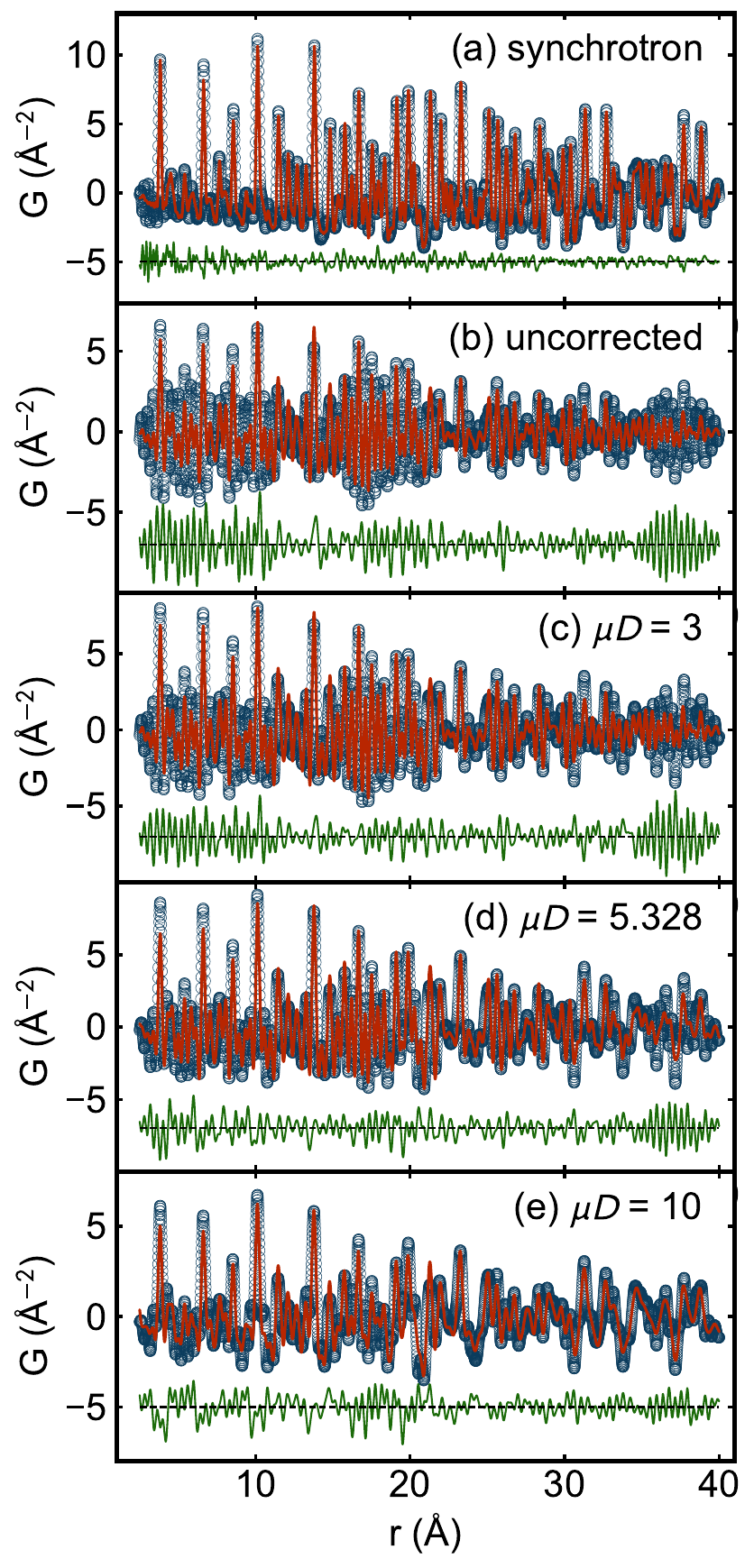}
    \caption{\ce{CeO2} PDF fits for synchrotron, uncorrected, and corrected data with different \mud's (reported in the panels), over $r_{\min} = 1.0$ and $r_{\max} = 40.0$. The plots show the measured PDFs (blue), the best-fit PDFs (red), and the residuals (green).}
    \label{fig:ceo2_pdfs}
\end{figure}

\end{document}